\documentclass[%
 aip,
rsi,%
 amsmath,amssymb,
 reprint,%
]{revtex4-2}

\usepackage{xcolor}
\usepackage{graphicx}
\usepackage{dcolumn}
\usepackage{bm}
\usepackage{nicefrac}
\newcommand{\n}[1]{\mathbf{#1}}

\begin{document}

\preprint{AIP/123-QED}

\title[Rodr\'{i}guez et al.]{Constructing the space of quasisymmetric stellarators}

\author{E. Rodr\'{i}guez}
 \altaffiliation[Email: ]{eduardo.rodriguez@ipp.mpg.de}
 \affiliation{ 
Max Planck Institute for Plasma Physics, Greifswald, Germany, 17491
}

\author{W. Sengupta}
 \altaffiliation[Email: ]{wricksg@gmail.com}
 \affiliation{ 
Department of Astrophysical Sciences, Princeton University, Princeton, NJ, 08543
}
\affiliation{%
Princeton Plasma Physics Laboratory, Princeton, NJ, 08540
}%

\author{A. Bhattacharjee}
 \altaffiliation[Email: ]{amitava@princeton.edu}
 \affiliation{ 
Department of Astrophysical Sciences, Princeton University, Princeton, NJ, 08543
}
\affiliation{%
Princeton Plasma Physics Laboratory, Princeton, NJ, 08540
}%

\date{\today}

\begin{abstract}
A simplified view of the space of optimised stellarators has the potential to guide and aid the design efforts of magnetic confinement configurations suitable for future fusion reactors. We present one such view for the class of quasisymmetric stellarators based on their approximate description near their centre (magnetic axis). The result is a space that captures existing designs and presents new ones, providing a common framework to study them. Such a simplified construction offers a basic topological approach, guided by certain theoretical and physical choices, which this paper presents in detail. 
\end{abstract}

\maketitle

\section{\label{sec:intro} Introduction} \label{sec:intro}

Designing magnetic configurations suitable for holding a thermonuclear plasma is central to fusion research. However, finding the appropriate shape of magnetic fields with the desired properties is challenging. There are primarily two reasons for this difficulty. First, one must select an appropriate set of objectives, imposed both by physics and technology, which may not be mutually compatible. Second, the potential parameter space for three-dimensional fields is very large.
\par
Most three-dimensional configurations are not attractive candidates, mainly because of their poor confinement properties. In an inhomogeneous magnetic field, charged particles generally drift away from field lines unless the field is carefully designed. The class of stellarators where charged particles are, on average, confined (collisionlessly), are called \textit{omnigeneous}\cite{bernardin1986,cary1997,hall1975,landreman2012,Helander2014}. This paper focuses on a subgroup of omnigeneous stellarators known as \textit{quasisymmetric} stellarators \cite{boozer1983,nuhren1988,rodriguez2020}. In these configurations, the magnitude of the magnetic field (but not the full vector field) has a direction of symmetry. The conventional approach to finding this particular subset of configurations remains an extensive search in the space of all possible stellarators. The search attempts to `minimise' the asymmetries in the magnetic field magnitude and has successfully provided multiple designs.\cite{Anderson1995,Zarnstorff2001,Najmabadi2008,Ku2010,bader2019} However, the optimisation procedure remains, to a large extent, a black box. This leaves important questions about quasisymmetry and its implications unanswered. Moreover, and perhaps most importantly, the black box can miss out on a significant number of undiscovered designs of potentially high value.  
\par
This paper presents an attempt to shed some light on these questions by considering an alternative view on the configuration space of quasisymmetric stellarators. This alternative view consists of a model based on an approximate description of stellarators close to their magnetic axis, where the complexity of the stellarator is significantly reduced. In such a framework, potential quasisymmetric stellarators are reduced to a combination of a magnetic axis shape and the choice of two scalar parameters, in terms of which many relevant properties can be expressed. Such a description provides the space with a basic topological structure that enables deeper understanding. Several choices are necessary for the axes of such a model to represent quasisymmetric stellarators. This paper presents the formal and physical basis underlying such choices, as well as the resulting highlights of the approach.
\par
Section II introduces the basics of the truncated near-axis expansion that constitutes the basic quasisymmetric stellarator model used. Sections III-V take the elements of such a model and explore their physical implications, guiding the appropriate choice of parameters for the model. The focus here will not be on the axis shapes, on which a thorough discussion may be found elsewhere\cite{rodriguez2022phase}, but rather on choosing the remaining parameters that complete the model. Section VI then presents an example of QS configuration space, which we analyse to illustrate the potential of this approach. We conclude with some final remarks and open questions.

\section{Quasisymmetry and near-axis expansions} \label{sec:introQS}
We begin by introducing the notion of \textit{quasisymmetry}(QS). This hidden symmetry is the minimal property of a magnetic field that provides the dynamics of charged particles with an approximate conserved dynamical quantity (\textit{Tamm's theorem}) to leading order in the gyroradius.\footnote{We simplify the picture by assuming that the electrostatic potential shares the QS to leading gyro-order and do not include it in our considerations.} Such a conserved momentum prevents particles from freely escaping the magnetic field, making the concept naturally attractive for magnetic confinement.\cite{tessarotto1996,burby2020,rodriguez2020} The condition can be formally expressed as $\nabla\psi\times\nabla B\cdot\nabla(\mathbf{B}\cdot\nabla B)=0$, where $2\pi\psi$ is the toroidal flux and $\mathbf{B}$ is the magnetic field. However, this form hides the underlying nature of QS, which is to make the contours of $|\mathbf{B}|$ symmetric. Under the assumption of ideal magnetohydrostatic equilibrium, $\mathbf{j}\times\mathbf{B}=\nabla p$, where $\mathbf{j}$ is the current density, in Boozer coordinates\cite{boozer1983}, QS implies\cite{rodriguez2021opt} that $|\mathbf{B}|=B(\psi,\chi=\theta-N\phi)$ is a function that depends on a linear combination of Boozer angles.\footnote{For a more general form of equilibrium, a formally analogous approach exists in terms of so-called generalised Boozer coordinates, details of which may be found in [\onlinecite{rodrigGBC}]. As a result, many of the properties of quasisymmetric stellarators in this paper are independent of a particular form of equilibrium.} Here, $N\in\mathbb{Z}$ describes the pitch of the symmetry, which leads to the distinction between quasi-axisymmetric (QA, $N=0$) and quasi-helically symmetric (QH, $N\neq0$) stellarators.
\par
Our goal is to describe stellarators with the property of QS (in the second form discussed above) close to the magnetic axis. The magnetic axis is the centre of the stellarator, a closed magnetic field line around which magnetic flux surfaces accrue. Because we are considering a description of the stellarator near its axis, it is natural to use the axis as reference for our coordinate system. We describe flux surfaces (i.e., constant $\psi$ surfaces, which we assume to be nested) using the Frenet-Serret basis\cite{garrenboozer1991a,landreman2019,rodriguez2020i} $\{\hat{b},\hat{\kappa},\hat{\tau}\}$ and Boozer coordinates $\{\psi,\theta,\phi\}$, so that with respect to the axis $\mathbf{r}_0$, we write
\begin{equation}
    \mathbf{x}=\mathbf{r}_0+X(\psi,\theta,\phi)\hat{\kappa}+Y(\psi,\theta,\phi)\hat{\tau}+Z(\psi,\theta,\phi)\hat{b}, \label{eqn:xDef}
\end{equation}
where $X,~Y$, and $Z$ are functions of all Boozer coordinates. This paper uses the notation in [\onlinecite{rodriguez2020i}], including the convention on the sign of the torsion.
Not every flux surface shape described by Equation ~(\ref{eqn:xDef}) is consistent with a given divergenceless magnetic field, which must be in equilibrium and be quasisymmetric. Let us start by imposing the condition that the magnetic field is solenoidal ($\nabla\cdot\mathbf{B}=0$) and lives on flux surfaces ($\mathbf{B}\cdot\nabla \psi=0$) formally\cite{kruskuls1958} by writing both the covariant and contravariant forms of $\mathbf{B}$ and using Boozer coordinates as independent coordinates, \cite{garrenboozer1991a,rodriguez2020i,rodrigGBC},
\begin{align}
    (B_\alpha(\psi)-&\Bar{\iota}B_\theta)\frac{\partial\n{x}}{\partial\psi}\times\frac{\partial\n{x}}{\partial\chi}+B_\theta\frac{\partial\n{x}}{\partial\phi}\times\frac{\partial\n{x}}{\partial\psi}+B_\psi\frac{\partial\n{x}}{\partial\chi}\times\frac{\partial\n{x}}{\partial\phi}=\nonumber\\
    &=\frac{\partial\n{x}}{\partial\phi}+\Bar{\iota}(\psi)\frac{\partial\n{x}}{\partial\chi}, \label{eq:co(ntra)variant}
\end{align}
where $\mathbf{x}$ is defined in Eq.~(\ref{eqn:xDef}). Here $\bar{\iota}=\iota-N$ and $\iota$ is the rotational transform, and $B_i$ are the covariant components of the magnetic field.
\par
 In Boozer coordinates, a quasisymmetric magnetic field satisfies
\begin{equation}
    \frac{B_\alpha(\psi)^2}{B(\psi,\chi)^2}=\left|\frac{\partial\n{x}}{\partial \phi}+\Bar{\iota}\frac{\partial\n{x}}{\partial \chi}\right|^2. \label{eqn:Jgen}
\end{equation} 
As given, Eqs.~(\ref{eq:co(ntra)variant}) and (\ref{eqn:Jgen}) constitute a coupled set of partial differential equations (PDEs) describing a quasisymmetric magnetic field, not only near the magnetic axis but everywhere. Unless we consider them close to the axis, the system of equations is overly complicated. 
\par
The asymptotic description of the fields near the magnetic axis is known as the \textit{near-axis expansion}\cite{Mercier1964, Solovev1970}, in the form here presented pioneered by [\onlinecite{garrenboozer1991a}]. This procedure entails expanding all relevant fields in the problem as power series in the distance from the magnetic axis. A pseudo-radial coordinate $\epsilon=\sqrt{\psi}$ is defined, where $2\pi\psi$ is the toroidal flux, which serves as the appropriate ordering parameter.\footnote{For simplicity, we have not normalised $\psi$ respect to the magnetic field on the axis and its curvature as it is often customary\cite{garrenboozer1991a,landreman2019}. Doing so simply introduces constant rescaling factors in various quantities involved.} Because of its radial nature, the expansion in $\epsilon$ must be carefully coupled to the poloidal-angle, $\theta$, behaviour. To avoid a coordinate singularity on the magnetic axis, all physical quantities must have the following asymptotic form,
\begin{equation}
    f=\sum_{n=0}^\infty\epsilon^n{\sum_{m=0|1}^{n}}\left[f_{nm}^c(\phi)\cos m\chi+f_{nm}^s(\phi)\sin m\chi\right]. \label{eqn:FourTayExp}
\end{equation} 
If the function $f$ is a flux function, the expansion reduces to a Taylor expansion in $\epsilon^2$. In this paper, we will use this subscript notation repeatedly.
\par
Expanding all functions in Eqs.~(\ref{eq:co(ntra)variant})-(\ref{eqn:Jgen}) as the coupled Fourier (in $\chi$) -Taylor (in $\epsilon$) series described by Eq.~(\ref{eqn:FourTayExp}), the PDEs are reduced into an (a priori) infinite ordered set of ordinary differential and algebraic equations. By order $n$, we are referring here to all the elements in the problem that have the same power $\epsilon^n$. At each order, various parameters and functions are needed as inputs to the equations to uniquely determine the solution, as summarised in Table \ref{tab:QSconfigCharParam}. The main elements are the magnetic field on the axis ($B_0$) and its leading variations nearby ($\eta,~B_{22}^C$ and $B_{22}^S$), the shape of the axis, the leading contribution of the toroidal current ($B_{\theta20}$), the stellarator-symmetry breaking ($\sigma(0)$) and the pressure gradient ($B_{\alpha2}$). Each choice represents a different stellarator; in that sense, the framework described here serves as a reduced stellarator model. We do not present the detailed order-by-order set of equations that constitute the near-axis description, as these may be found elsewhere, both in the form that concerns us here\cite{landreman2019,garrenboozer1991b} and in the context of more general equilibria\cite{rodriguez2021weak, thesis}. Instead, we focus on the choice of parameters that make the near-axis model represent optimised quasisymmetric configurations.
\par
\begin{table}[]
    \centering 
    \begin{tabular}{|c|c|}
        \hline
    Order & Params. \\ \hline\hline
    $0$ & $B_0$, axis ($\kappa,~\tau,~l)$ \\
    $1$ & $B_{\theta 20}$, $\sigma(0)$, $\eta$ \\
    $2$ & $B_{22}^C$, $B_{22}^S$, $B_{\alpha2}$ \\\hline
    \end{tabular}
    \caption{\textbf{Quasisymmetric configuration characterising parameters.} The table gathers the free parameters (and functions) defining the leading order form of quasisymmetric configurations.}
    \label{tab:QSconfigCharParam}
\end{table}

\section{Zeroth order: magnetic axis} \label{sec:0th}
Let us start with the most basic element in the model: the shape of the magnetic axis (see Tab.~\ref{tab:QSconfigCharParam}). At a fundamental level, the near-axis model identifies every configuration with a three-dimensional closed curve (magnetic axis). Every configuration that shares the same magnetic axis must then also share certain properties. A detailed discussion on the role of the magnetic axis in the context of quasisymmetric stellarators was presented in [\onlinecite{rodriguez2022phase}]. Here we reproduce the essential elements concerned in constructing our quasisymmetric stellarator model.
\par
From the set of all smooth, three-dimensional closed curves, those with inflection points, that is, points of vanishing curvature, must be excluded. To see why this is the case, interpret the curvature of the axis, $\kappa$, as a measure of the scale of the magnetic field magnitude gradient near the axis (from equilibrium $\nabla_\perp(B^2/2)=B^2\vec{\kappa}$). Thus, to support any finite magnetic field variation on a flux surface around a point where $\kappa\approx0$, an unphysical, nearly infinitely elongated flux surface is necessary\cite{landreman2018a,rodriguez2022phase}. Under such conditions, the set of excluded axis shapes becomes physically interesting only outside the rigorous realm of QS, as in the case of \textit{quasi-isodynamic} stellarators\cite{hall1975, plunk2019, rodriguez2023qi}. In QS, which is the focus of the present paper, the requirement of specialising to regular curves makes the Frenet-Serret frame well-defined everywhere.
\par
With such a frame defined, we can construct a \textit{self-linking number}\cite{rodriguez2022phase,oberti2016,aicardi2000,fuller1999, moffatt1992} ($S_L$), which is the number of times the curvature vector of the axis encircles itself in a full toroidal excursion. This number is \textit{precisely} the integer $N\in\mathbb{Z}$ that appears in the quasisymmetric form of $|\mathbf{B}|=B(\psi,\theta-N\phi)$. Thus, the shape of the axis (a local feature) fully determines the class of the quasisymmetric stellarator (a global feature)\cite{rodriguez2022phase,landreman2018a}. 
\par
Besides its global implications, the strength of this association is that $S_L$ is a topological invariant under regular isotopies. The property remains unchanged if the magnetic axis is continuously deformed provided that the curvature does not vanish anywhere along the way. Thus, the space of all closed curves is partitioned into regions identified by an integer value of $S_L$.
Each quasisymmetry class thus behaves as a \textit{quasisymmetric phase}, whose nature may only be changed by crossing the separating manifolds, \textit{phase-transitions}, made up of curves with inflection points. It is then natural to see each phase as a distinct class. 
\par
Lacking a theory that directly relates axis shapes to quasisymmetric quality of the stellarator, we must consider all (regular) axis shapes as part of our space of potential QS configurations. Elements from higher orders in the near-axis expansion will be needed to tell different shapes apart and delve into additional properties of the configurations. It is important to note the difference between this model space and the one more traditionally employed in optimisation: the space of toroidal surface shapes. The consideration of the axis reduces the dimensionality of the space and, crucially, provides such a space with topological structure [\onlinecite{rodriguez2022phase}]. It is evident from the latter that an optimisation that starts in a particular phase will remain within it. 
\par
There is no unique way to represent this space of closed curves, and we consider a Fourier description of the curves in cylindrical coordinates $(R(\phi), Z(\phi))$ for simplicity. Here, $\phi$ is the cylindrical angle, and $R=\sum_n R_n\cos nN\phi$ and $Z=\sum_n Z_n\sin nN\phi$. We specialise to stellarator-symmetric configurations. This parametrisation of the curves guarantees they are closed and have $N-$fold symmetry. However, the torsion, $\tau$, and curvature, $\kappa$, which are most directly involved in the near-axis construction become byproducts that (as is the case of torsion) can be pretty sensitive to the choice of Fourier harmonics. Parametrising the curves providing $\kappa$ and $\tau$, which we may call the \textit{Frenet approach}, would be more natural (and in agreement with the fundamental theorem of curves\cite{fenchel1951}) but suffer from the issue of closing the curve. Other possible alternatives, such as the use of control points of splines, have generated renewed interest\cite{paz2020,lonigro2022}. The story of representations is mixed: the advantages of one appears to be the Achilles heel of the other. However, the foundations of this paper are independent of the particular form of a representation.

\section{First order: elliptic shaping} \label{sec:1st}
With the magnetic axis in place, let us proceed to the first order. Following Table~\ref{tab:QSconfigCharParam}, we must choose appropriate values for three parameters: the toroidal current $B_{\theta 20}$, the stellarator-symmetry breaking $\sigma(0)$, and the $|\mathbf{B}|$ variation, $\eta$. Different choices will describe different configurations, thereby lifting part of the degeneracy that results from identifying stellarators with their magnetic axis. In this order, the model introduces elements of flux surface shaping and the rotational transform. We will see how the choice of parameters in this order affects the behaviour of these features, and provide guidelines for appropriate choices. 
\par

\subsubsection{Toroidal current, $B_{\theta20}$}
The coefficient $B_{\theta20}$ controls the plasma current density on the axis and is, in more familiar notation, the leading contribution to $B_{\theta}=I(\psi)$. In constructing most QS configurations, the trivial assumption $B_{\theta 20}=0$ is often made. We shall do so here as well. Of course, such a choice is exact when modelling a vacuum field, but it is not always the appropriate limit for configurations that support a finite plasma pressure. Plasma currents, such as bootstrap currents, may be present in that case, even without having to drive them externally. Their evaluation generally requires a separate kinetic consideration \cite{helander2005,ware2006}. Bearing this caveat in mind, we specialise to vacuum magnetic fields and thus take $B_{\theta20}=0$. The problem does not generally show singular behaviour in this limit. Thus, this simplifying assumption is justified for illustrating the approach in this paper.  

\subsubsection{Surface shaping and rotational transform: $\eta$ and $\sigma(0)$}
We now consider the relevant parameters $\eta$ and $\sigma(0)$. By definition, the parameter $\eta=-B_{11}^C/2B_0$ is a measure of the variation of the magnetic field over flux surfaces. The other parameter, $\sigma(0)$, is $\sigma=Y_{11}^C/Y_{11}^S$, a quantity related to the shaping of flux surfaces at the point of stellarator symmetry $\phi=0$. Beyond these definitions, both of these parameters may be connected to natural elements of the geometry of flux surfaces (see [\onlinecite{rodriguez2023mhd}] for more details). The shapes of the flux surfaces at the first order are elliptical and can be characterised on the plane orthogonal to the magnetic axis by 
an elongation $\mathcal{E}$, and rotation angle, $\vartheta$,
\begin{subequations}
\begin{gather}
    \mathcal{E}=\frac{F}{\eta^2/\kappa^2}\left[1+\sqrt{1-\frac{\eta^4/\kappa^4}{ F^2}}\right],
    \label{eqn:aspRatioExpress} \\
    \tan 2\vartheta=\frac{\sigma\eta^2/\kappa^2}{\eta^4/4\kappa^4-1-\sigma^2}, \label{eqn:rotAngleEllip}
\end{gather}
\end{subequations}
where $F=1+\sigma^2+\eta^4/4\kappa^4$. Here \textit{elongation} is defined as the ratio of the major to the minor radius, and $\vartheta$ is the angle between the major radius and the positive $X$ direction (i.e., the normal vector).
\par
From Eq.~(\ref{eqn:rotAngleEllip}), it is clear that the parameter $\sigma(0)$ serves as a measure of the misalignment of the ellipse with the Frenet-Serret frame at the origin, $\phi=0$. That is, it is a measure of the up-down asymmetry of this cross-section and, generally, of stellarator asymmetry. If we specialise to the stellarator-symmetric choice ($\sigma(0)=0$), a single parameter is then left to choose, $\eta$.
\par
From the form of $\mathcal{E}$, Eq.~(\ref{eqn:aspRatioExpress}), in the limit of $\sigma=0$, where $\mathcal{E}\sim\eta^2/2\kappa^2,~2\kappa^2/\eta^2$ (the latter for $\eta<\kappa\sqrt{2}$), we may interpret $\eta$ as an approximate measure of elongation. A large $\eta$ indicates a large elongation in the direction of the curvature vector, while a small $\eta$ corresponds to large elongation along the binormal. This correspondence is only approximate because even with the stellarator-symmetric choice $\sigma(0)=0$, $\sigma$ is generally a non-zero function of $\phi$, which makes Eq.~(\ref{eqn:aspRatioExpress}) highly non-trivial. 
\par
\begin{figure}
    \centering
    \includegraphics[width=0.45\textwidth]{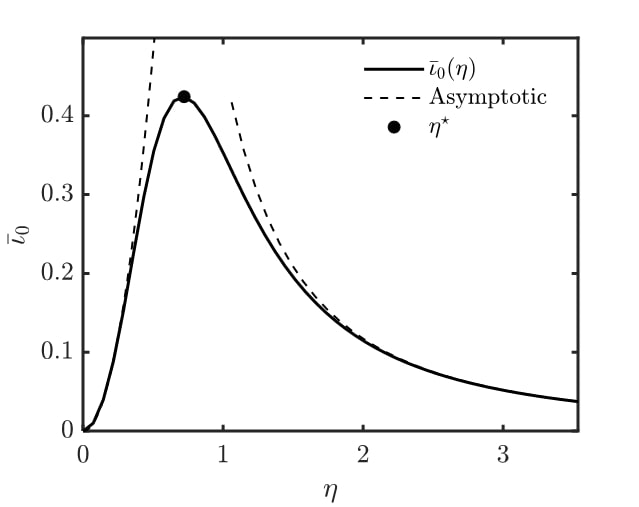}
    \caption{\textbf{Example of $\bar{\iota}_0$ as a function of $\eta$.} Example of the rotational transform as a function of $\eta$ for a fixed axis shape. This example corresponds to a quasiaxisymmetric configuration, with an axis like the precise QA in [\onlinecite{landreman2022}]. The broken lines indicate the asymptotic behaviour described in the text.}
    \label{fig:my_label}
\end{figure}

This naturally takes us to the equation that governs the form of $\sigma$, a first-order non-linear differential equation, 
\begin{equation}
    \frac{\mathrm{d}\sigma}{\mathrm{d}\phi}=-\bar{\iota}_0\left[1+\sigma^2-\frac{1}{4B_0}\left(\frac{\eta}{\kappa}\right)^4\right]+\frac{B_{\alpha0}}{2}(2\tau+B_{\theta20})\left(\frac{\eta}{\kappa}\right)^2, \label{eqn:sigmaEqn}
\end{equation}
referred to as the $\sigma$ Riccati equation\cite{garrenboozer1991b,landreman2019,rodriguez2021weak}. The choice of $\eta$ affects $\sigma$ non-trivially, and thus the shaping of the flux surfaces. Following Mercier\cite{Mercier1964}, we then expect the rotational transform of the configuration on the axis, $\iota_0$ (or more generally in quasisymmetry, $\bar{\iota}_0=\iota-N$, where $N$ is the self-linking number of the axis) to depend on $\eta$ as well. Formally, $\bar{\iota}_0$ forms part of the solution to the Riccati equation, Eq.~(\ref{eqn:sigmaEqn}), as there is a unique value for which $\sigma(\phi)$ is periodic\cite{landreman2019}. Thus, through the lens of this equation, we may regard the rotational transform as a function of $\eta$ for a fixed axis shape (see Figure~\ref{fig:my_label}).
\par
 No closed form expression can generally be obtained for $\bar{\iota}_0(\eta)$. However, the behaviour of $\bar{\iota}_0(\eta)$ for large and small values of $\eta$ can be determined. In the small $\eta$ limit, the dominant balance $\sigma\sim\eta^2$ and $\bar{\iota}_0\sim\eta^2$ of Eq.~(\ref{eqn:sigmaEqn}) yields, upon integration,
\begin{equation}
    \bar{\iota}_0\sim\bar{\iota}_-\eta^2=\frac{B_{\alpha0}}{4\pi}\eta^2\int_0^{2\pi}\frac{2\tau+B_{\theta20}}{\kappa^2}\mathrm{d}\phi. \label{eqn:asymSmallEtaIota}
\end{equation}
Thus, the rotational transform on the axis tends to $\iota_0\rightarrow N$. In the case of a quasi-axisymmetric (QA) configuration, this leads to a vanishing rotational transform. Physically, this is a result of making flux surfaces very elongated in the curvature direction in a way that, from Mercier's perspective, the rotational transform on the axis is driven only by the rotating-ellipse contribution. Of course, in the QA case, the ellipse has no net rotation (recall the meaning of the self-linking number).\footnote{From the Mercier perspective on rotational transform\cite{Helander2014},
$$     \bar{\iota}_0=-\frac{1}{2\pi}\int_0^L\frac{\cosh\bar{\eta}-1}{\cosh\bar{\eta}}(d'+\tau)\mathrm{d}l+\frac{1}{2\pi}\int_0^L\tau\mathrm{d}l, \label{eqn:MercierIota}
 $$ in the large ellipticity $\bar{\eta}\rightarrow\infty$ limit, $\bar{\iota}_0=-d/2\pi$, where $d$ is the angle of rotation of the ellipse with respect to the curvature vector. Because $\sigma\sim\eta^2\rightarrow0$ in this limit, the cross-sections align with the Frenet-Serret frame. Thus, the net rotation $d=0$.} Thus, $\iota_0=0$. As $\eta$ increases, the torsion and current contributions play the same role in driving rotational transform. In a QA configuration, this growth in $\bar{\iota}_0$ is equivalent to growth in rotational transform. However, in the case of a quasi-helically symmetric (QH) configuration, this depends on the relative sign of Eq.~(\ref{eqn:asymSmallEtaIota}) and the QS helicity $N$, which tends, in practice (and zero current), to be negative. In the limit of large $\eta$, the physical scenario is similar to that in the small $\eta$ limit. The shaping becomes large, and $\bar{\iota}_0$ also tends to zero. This follows formally from the dominant balance $\bar{\iota}_0=\bar{\iota}_0^*/\eta^2$ and $\sigma=\eta^2\sigma^*$ in Eq.~(\ref{eqn:sigmaEqn}). The asymptotic forms of $\{\sigma^*,\bar{\iota}_0^*\}$ are solutions to a `universal' $\eta$-independent Riccati equation, unique to each axis shape ($\kappa,~\tau$). 
\par
By the mean value theorem, it follows that $\iota_0(\eta)$ must have at least one turning point (except in the marginal case of $\int(2\tau+B_{\theta20})/\kappa^2=0$). In Appendix~\ref{sec:appEtaStar} we prove that this extremum exists, its value $\eta=\eta^*$ is unique and satisfies the condition,
\begin{equation}
        \int_0^{2\pi}\left[\sigma^2+\frac{1}{4B_0}\left(\frac{\eta}{\kappa}\right)^4-1\right]E\mathrm{d}\varphi=0, \label{eqn:extrEtaIotaRel}
\end{equation}
where $E=\exp[{2\bar{\iota}_0\int_0^{\varphi}\sigma\mathrm{d}\varphi'}]$. From this expression, it follows that there is a value of $\eta$ in $(0,\sqrt{2}\kappa_\mathrm{max}B_0^{1/4}]$ that extremises the rotational transform (see Fig.~\ref{fig:my_label}). 
\par
To illustrate further the meaning of $\eta^*$, and to further connect it to shaping, consider the familiar limit of axisymmetry. In that case, the Riccati equation, Eq.~(\ref{eqn:sigmaEqn}), can be solved exactly (taking $B_{\alpha0}=1=R_0$) to yield
\begin{equation}
    \iota_0=\frac{1}{2}\frac{B_{\theta20}\eta^2}{1+\eta^4/4}. \label{eqn:ASiota}
\end{equation} 
The solution in Eq.~(\ref{eqn:ASiota}) exhibits all the properties of $\bar{\iota}_0$ we studied in the general case, including the uniqueness of $\eta^*=\sqrt{2}$. This choice of $\eta$ corresponds to having a circular cross-scetion. This suggests that in order to maximise the use of the toroidal current to generate rotational transform, one must simultaneously minimise the amount of shaping. 
\par 
The correspondence between the behaviour of the rotational transform and the shaping of surfaces prevails beyond axisymmetry. This can be seen by investigating the behaviour of elongation $\mathcal{E}$, Eq.~(\ref{eqn:aspRatioExpress}), with $\eta$ in an analogous form to the case of $\bar{\iota}_0$. Its asymptotic behaviour was already touched upon before, as we saw that elongation diverged both for small $\eta$ ($\mathcal{E}\approx2\sqrt{B_0}(\kappa/\eta)^2$) and for large $\eta$ ($\mathcal{E}\sim2\eta^2\sqrt{B_0}\kappa^2[\tilde{\sigma}^2+1/4B_0\kappa^4]$). Thus, there must be some minimally elongated configuration somewhere in between.
\begin{figure}
    \centering
    \includegraphics[width=0.5\textwidth]{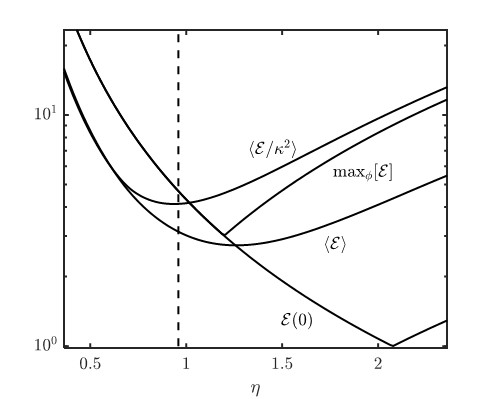}
    \caption{\textbf{Difference in behaviour between elongation measures.} Plot showing the behaviour of different elongation measures with $\eta$ for the magnetic axis of the `precise QA' configuration\cite{landreman2022}. The plot shows the difference in behaviour between different global scalar measures for elongation: $\mathcal{E}(0)$ is the value of elongation at $\varphi=0$, $\langle\mathcal{E}\rangle$ is the $\phi$-average of elongation, $\mathrm{max}_\varphi[\mathcal{E}]$ is the maximum of elongation and $\langle\mathcal{E}/\kappa^2\rangle$ is the weighted average of elongation. The broken line represents the value of $\eta$ obtained from the global solution for comparison.}
    \label{fig:ellonPlot}
\end{figure}
However, this notion must be qualified, as $\mathcal{E}$ is not a single scalar but a function of $\varphi$. This allows for different ways in which to construct a single measure for the shaping of the configuration. Figure \ref{fig:ellonPlot} shows some possibilities. It is evident that the behaviour changes from one definition to another. Among these measures, there is one that is particularly insightful. Consider the weighted average $\bar{\mathcal{E}}=(1/2\pi)\int(\mathcal{E}/\kappa^2)\mathrm{d}\phi$ which penalises elongation in the straighter sections of the configuration. This particular elongation measure has the same asymptotic behaviour (up to an $\eta-$independent factor) as $\bar{\iota}_0$. This formal analogy makes the choice of $\eta^*$ also a choice that roughly minimises elongation (see the comparison in Fig.~\ref{fig:gradBmeasureothers} or Table~\ref{tab:compEtadesigns} for the implications in practice).  
\par
Besides all the physical implications of $\eta^*$, this choice also represents a least sensitive choice, so much so that it corresponds to a turning point. That is, a set of near-axis constructions in some range $(\eta^*-\delta,\eta^*+\delta)$ have roughly the same properties, making their properties more robust. Following [\onlinecite{rodriguez2021weak}], this resilience of the rotational transform makes the magnetic shear (a quantity that would be, in general, a third-order quantity) independent of third-order parameter choices (see Appendix~\ref{sec:appShear}). Thus, our model can predict the behaviour of magnetic shear as well. 
\par
\begin{table*}
    \centering
    \begin{tabular}{c|cccccccc}
        & ARIESCS & ESTELL & GAR & HSX & NCSX & QHS48 & Precise QA & Precise QH \\\hline
        $\eta_\mathrm{VMEC}$ & 0.11 & 0.79 & 0.51 & 1.75 & 0.62 & 0.21 & 0.96 & 2.12 \\
        $\eta_\mathrm{VMEC}/\eta^*$ & 1.00 & 0.91 & 0.95 & 0.87 & 0.92 & 0.88 & 0.95 & 0.91 \\
        $\eta_\mathrm{VMEC}/\eta_{\langle\nabla \mathbf{B}\rangle}$ & 1.03 & 0.12 & 0.57 & 0.99 & 0.73 & 1.07 & 0.64 & 1.04\\
        $\eta_\mathrm{VMEC}/\eta_{\bar{\mathcal{E}}}$ & 1.07 & 0.95 & 1.01 & 0.91 & 0.97 & 0.91 & 1.03 & 0.95 \\
        $N$ & 0 & 0 & 0 & 4 & 0 & 4 & 0 & 4 \\\hline
    \end{tabular}
    \caption{\textbf{Comparison of $\eta$ choices in some quasisymmetric designs.} Comparison of the $\eta$ values for many quasisymmetric designs: ARIESCS\cite{Najmabadi2008}, ESTELL\cite{Drevlak2013}, GAR\cite{Garabedian2008,Garabedian2009}, HSX\cite{Anderson1995}, NCSX\cite{Zarnstorff2001}, QHS48\cite{Ku2010}, Precise QA and QH\cite{landreman2022}. $\eta_\mathrm{VMEC}$ corresponds to the value of $\eta$ obtained from the $|\mathbf{B}|$ of the global \texttt{VMEC}\cite{hirshman1983} equilibrium solution of the quasisymmetric designs. $\eta^*$ is the parameter value that extremises $\bar{\iota}$ for a fixed axis obtained from \texttt{VMEC}. $\eta_{\langle\nabla\mathbf{B}\rangle}$ corresponds to the maxima of the average of $L_{\nabla\mathbf{B}}$ (see Appendix~\ref{sec:appGradB}), and finally $\eta_{\bar{\mathcal{E}}}$ from the minimum of the weighted average elongation $\bar{\mathcal{E}}$. The integer $N$ denotes the symmetry class. }
    \label{tab:compEtadesigns}
\end{table*}
In summary, the choice of $\eta=\eta^*$ is a formally convenient and representative choice for $\eta$. It is unique and always exists, maximises rotational transform in QA configurations, and regularises the shaping of flux surfaces, preventing them from having extreme shaping. With such a choice, every axis in our structured space of configurations will have an associated natural choice of $\eta$. We do not need to keep this parameter explicitly as an added dimension in this space, reducing its complexity. A practical design would benefit from an additional tweaking of this parameter in the neighbourhood of $\eta^*$. This refinement could be seen as a subsequent optimisation in which one can consider higher-order properties or more sophisticated construction specifications.
\par
To present some evidence that backs the suitability of the choice of $\eta^*$, we present in Table~\ref{tab:compEtadesigns} the value of $\eta$ extracted from global equilibrium solutions designed through other means compared to different choices of $\eta$. This shows that $\eta^*$ is a reasonable representative choice across the board. In the Table, we also include a choice $\eta_{\nabla\mathbf{B}}$, which we have not mentioned in the discussion. This choice of $\eta$ is taken to be the value that minimises the magnitude of $||\nabla\mathbf{B}||$, a measure of the gradients of the magnetic field. This measure has been used by other researchers\cite{landreman2021a,landreman2022,landreman2022map} to good effect in optimising near-axis QS configurations. The rationale behind this measure is that the characteristic length scale $L_\nabla\stackrel{\cdot}{=}1/||\nabla\n{B}||$, may be interpreted to approximately set the distance to the field-generating coils, and thus to indicate a measure of the range of validity of the near-axis model.\cite{landreman2021a} Guiding the choice of $\eta$ for a fixed axis could be considered as a guiding principle in choosing $\eta$. This is successful in many cases (see Tab.~\ref{tab:compEtadesigns}), but it lacks the robustness and generality of $\eta^*$, and fails in the QA phase. See Appendix~\ref{sec:appGradB} for a more detailed discussion.
\par

\section{Second order: surface triangularity and QS breaking} \label{sec:2nd}
So far, we have identified quasisymmetric configurations with a model consisting of an axis and elliptical cross-sections. For every curve in our space of configurations, the shape of the latter results from the choice of the parameter $\eta^*$. However, we note that this model is exactly quasisymmetric up to this point, and that there is no way of differentiating which configurations will exhibit better QS globally. Consideration of the second order in the near-axis expansion is needed for this. 
\par
When incorporating the second order, two important parameter choices must be made: (i) the plasma pressure gradient (in the notation here, related directly to $B_{\alpha1}$) and (ii) the second order variation of the magnetic field magnitude, $B_2$. For every choice made, the model gains a different flux-surface shaping in the form of triangularity, Shafranov shift, and a different degree of QS. In line with the simplifying vacuum and stellarator-symmetric assumptions, we choose $B_{\alpha1}=0$ and $B_{22}^S=0$ to reduce the number of free parameter choices to one. Only one of the harmonics of $B_2$ remains. Following the approach of [\onlinecite{landreman2019}], this parameter is taken to be $B_{22}^C$. This leaves the remaining component of $B$ at second order, $B_{20}$, to be found self-consistently. This lack of freedom in $B_{20}$ follows from the necessity of satisfying force-balance and the appropriate magnetic equations simultaneously, which generally requires $B_{20}$ to have a toroidal angle $\varphi$ dependence.\cite{garrenboozer1991b} Of course, such a variation violates QS, and the variation of $B_{20}$, which we may define as $\Delta B_{20}$ bottom-to-peak, becomes a measure of QS quality at this order.
\par
Formally, $B_{20}$ is the solution to a second-order, linear differential equation. The $B_{20}$ equation may be found in the appendix of [\onlinecite{thesis}], and may be written as,
\begin{equation}
    \mathcal{A}\left(\frac{B_{20}}{B_0}\right)''+\mathcal{B}\left(\frac{B_{20}}{B_0}\right)'+\mathcal{C}\frac{B_{20}}{B_0}+\mathcal{D}=0, \label{eqn:B20Eq}
\end{equation}
where 
\begin{widetext}
\begin{subequations}
\begin{gather}
    \mathcal{A}=-\frac{B_{\alpha0}\eta}{2\kappa^2\bar{\iota}_0l'}\left[1+\frac{4B_0\kappa^4}{\eta^4}(1+\sigma^2)\right], \label{eqn:AB20} \\
    \mathcal{B}=\frac{2B_{\alpha0}\eta}{\bar{\iota}_0 l'}\frac{\kappa'}{\kappa^3}-\frac{4l'\sigma}{\bar{\iota}_0\eta}\tau, \label{eqn:BB20} \\
    \mathcal{C}=-\frac{l'}{2B_{\alpha0}\eta^3\kappa^2}\left[\bar{\iota}_0\left(4\kappa^4(1+\sigma^2)-\frac{3\eta^4}{B_0}\right)+8B_{\alpha0}\eta^2\kappa^2\tau\right],
\end{gather}
\end{subequations}
\end{widetext}
and $\mathcal{D}$ is a complicated expression given in Appendix~D.2, Eq.~(D25d) of [\onlinecite{thesis}] (see also Appendix~\ref{sec:B20appendix}).
\par
Relaxing QS through $B_{20}$ and not other components of $|\mathbf{B}|$ is only a choice, not a requirement. It is, however, not a whimsical choice. For one, it has the benefit of making the self-consistent $B_{20}$ regular, in the sense of Fuchs criteria\cite{bender1999}. This follows from $\mathcal{A}\neq0$, as Eq.~(\ref{eqn:AB20}) is proportional to a sum of squares. This local consideration on $B_{20}$ is not to say that a solution to the equation that satisfies the condition of periodicity exists. Proving that consistent periodic solutions to Eq.~(\ref{eqn:B20Eq}) exist is a more challenging problem. To explore this question, let us first consider the axisymmetric limit.
\par
In the axisymmetric limit, everything is by definition $\varphi$-independent, and thus Equation~(\ref{eqn:B20Eq}) reduces (using the axisymmetric simplification of $[1+(1/4B_0)\eta^4/\kappa^4]\bar{\iota}_0=B_{\alpha0}B_{\theta20}\eta^2/2\kappa^2$) to,
\begin{equation}
    -\bar{\iota}_0^2\left(\eta^4-12B_0\kappa^4\right)\frac{B_{20}}{B_0}-\frac{3}{\eta^2\kappa^2}(\eta^4-4B_0\kappa^4)B_{22}^C=\dots. \label{eqn:B20AS}
\end{equation}
The dots denote second-order independent terms. Solving the equation for $B_{20}$ (the approach described above\cite{landreman2019}) clearly shows ill behaviour for $\eta^4=12\kappa^4B_0$, seemingly an arbitrary choice of elliptical shape. In that scenario, $B_{20}$ decouples from the equation, leaving only $B_{22}^C$ to satisfy the equation. The roles of $B_{20}$ and $B_{22}^C$ can be reversed to solve Eq.~(\ref{eqn:B20AS}) for $B_{22}^C$. However, in this case, the construction fails again at another value of $\eta$, i.e., whenever circular cross-sections are considered. To avoid excluding this case, solving for $B_{20}$ appears to be the more convenient choice.
\par
\begin{figure}
    \centering
    \hspace*{-0.5cm}
    \includegraphics[width=0.4\textwidth]{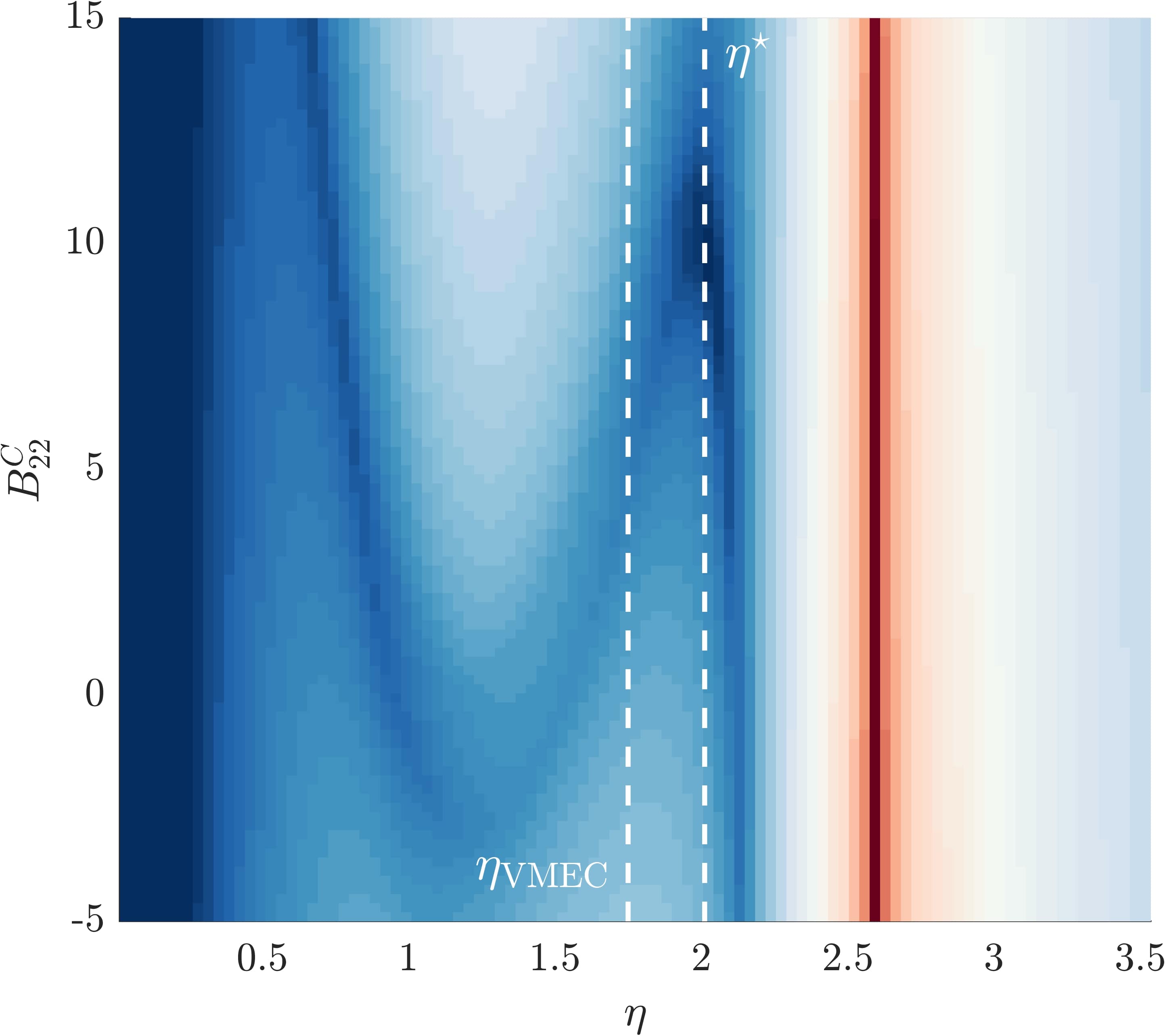}
    \caption{\textbf{Quasisymmetry breaking as a function of $\eta$ and $B_{22}^C$ for an example.} Variation in $B_{20}$ ($\Delta B_{20}$) as a function of the choice of parameters $\{\eta,B_{22}^C\}$ computed using the code \texttt{pyQSC}. The dashed lines correspond to the $\eta_\mathrm{VMEC}$ (from the global equilibrium) and $\eta^\star$ values (see Tab.~\ref{tab:compEtadesigns}). }
    \label{fig:singB20example}
\end{figure}

The breakdown of solutions for this `critical' value of $\eta$ persists beyond axisymmetry. Numerical evidence of this is presented in Fig.~\ref{fig:singB20example}. We plot $\Delta B_{20}$ for a fixed axis shape as a function of $\eta$ and $B_{22}^C$, which shows a clear critical value for $\eta$.
Formally, the appearance of such singularity may be explained through the \textit{Fredholm alternative theorem} for the existence of solutions. The singularity occurs when there is no solution to Eq.~(\ref{eqn:B20Eq}), which occurs whenever a solution exists to the adjoint problem. The adjoint problem does not generally have a solution (and thus, a solution to Eq.~(\ref{eqn:B20Eq}) exists and is unique), but one may rigorously prove that there is at least one critical value $\eta_\mathrm{crit}$ for which this is the case. If this critical value is unique (which numerical evidence suggests to be the case), it follows that $\eta^*<\eta_\mathrm{crit}$. Thus, our choice of $\eta^*$ at lower order guarantees the existence of a unique solution to Eq.~(\ref{eqn:B20Eq}). Details on this may be found in Appendix~\ref{sec:appB20Schrod}.
\par
\begin{figure}
    \centering
    \includegraphics[width=0.5\textwidth]{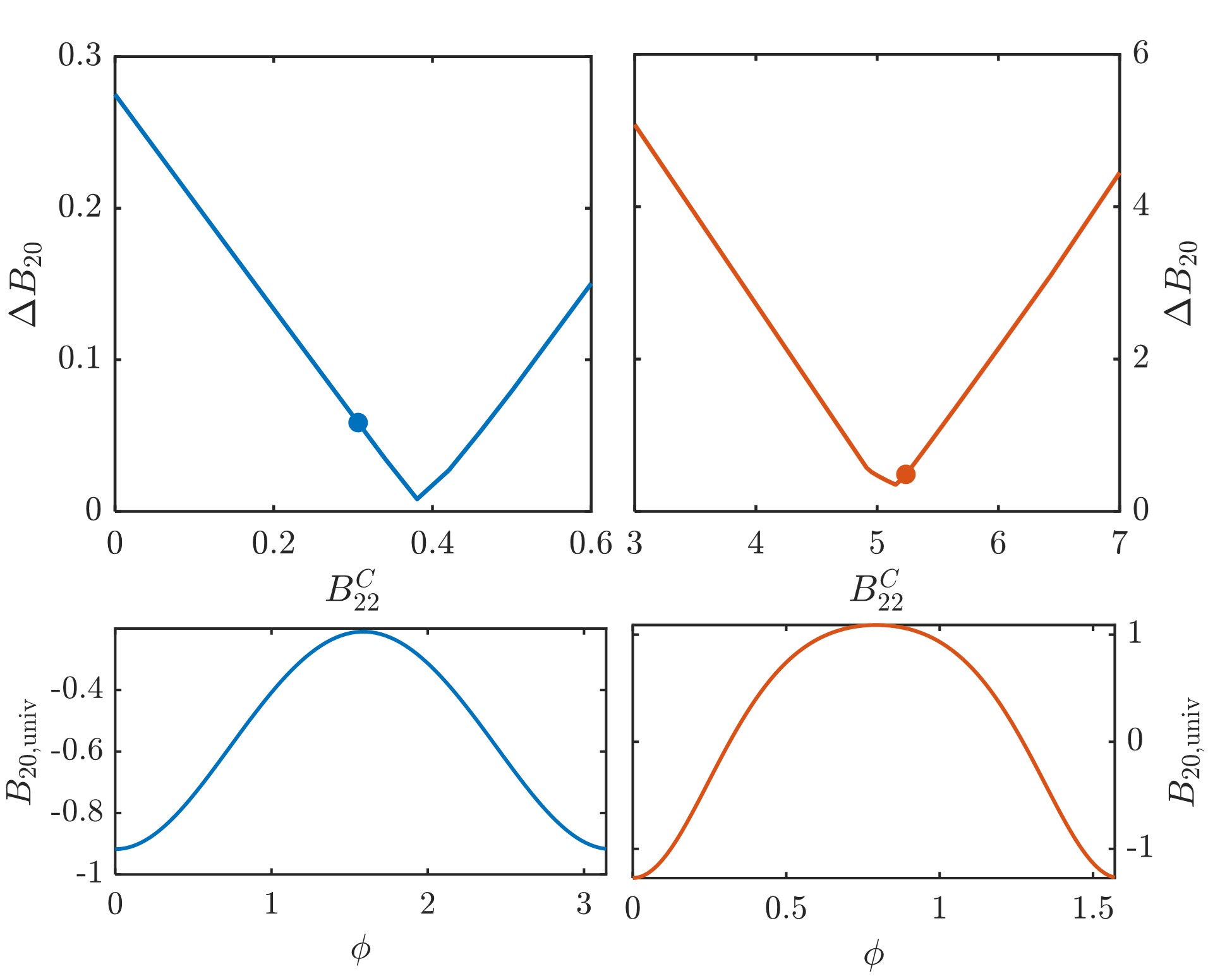}
    \caption{\textbf{Example of $B_{22}^C$ choice for precise QS designs.} Plots showing the variation of the QS residual as a function of $B_{22}^C$ for the NAE models of the precise QS designs \cite{landreman2022} (top) and their respective universal solutions $B_{20,\mathrm{univ}}$ (bottom). The plots correspond to the QA (left) and QH (right). The scatter points represent the values of $B_{22}^C$ obtained from the $|\mathbf{B}|$ spectrum of the global solutions. }
    \label{fig:B22cOptpreciseQS}
\end{figure}

The existence of a solution and this critical value $\eta_\mathrm{crit}$ have told us little about the influence of $B_{22}^C$, and how to choose it. However, the choice of $B_{22}^C$ follows naturally when we try to maximise the quality of QS. That is, we choose $B_{22}^C$ so that it minimises the deviation of $B_{20}(\varphi)$ from being a constant. This will make our model represent the `most' quasisymmetric configuration. Because the parameter $B_{22}^C$ appears only in the inhomogeneous term of Eq.~(\ref{eqn:B20Eq}), the choice of $B_{22}^C$ has no dramatic effect on $B_{20}$. Therefore, we expect the search of $B_{22}^C$ that minimises the QS residual to be smooth. More quantitatively, we write the dependence of $\mathcal{D}$, Eq.~(\ref{eqn:B20Eq}), on $B_{22}^C$,
\begin{multline*}
    \mathcal{D}=\frac{3\kappa^2}{B_{\alpha0}B_0(l')^2\eta^3}\left[\frac{\bar{\iota}_0}{2}\left(B_{\alpha0}^2\frac{\eta^4}{\kappa^4}+4(\sigma^2-1)\right)-\right.\\
    \left.-4\sigma'-8\bar{\iota}_0\sigma^2\right]B_{22}^C+\dots,
\end{multline*}
which in the limit of large $|B_{22}^C|$, makes $B_{20}$ scale with $B_{22}^C$, with a solution of the `universal' form,
\begin{equation}
    [\mathcal{A}\partial_\phi^2+\mathcal{B}\partial_\phi+\mathcal{C}]\frac{B_{20,\mathrm{univ}}}{B_0}+\mathcal{D}_C=0, \label{eqn:univB20}
\end{equation}
where $B_{20}=B_{22}^CB_{20,\mathrm{univ}}$. This is universal in the sense that each first-order construction has a single solution $B_{20,\mathrm{univ}}$. This solution provides a measure of the effect of $B_{22}^C$ on symmetry- breaking (see Figure~\ref{fig:B22cOptpreciseQS}). The influence of $B_{22}^C$ decreases as $B_{20,\mathrm{univ}}$ becomes closer to a constant to vanish for axisymmetry. Away from the neighbourhood of axisymmetry (where $B_{22}^C$ has little effect on QS), this means there must be at least one local minimum at a finite value of $B_{22}^C$. This makes minimising $\Delta B_{20}(B_{22}^C)$ modifying $B_{22}^C$ a well-posed 1D search problem. This way of choosing $B_{22}^C$ can be shown to be representative of optimised stellarators in practice (see Figure~\ref{fig:B22cOptpreciseQS}). 
\par
Although choosing $B_{22}^C$ this way constitutes a well-posed problem, we should not disregard other aspects of the stellarator that $B_{22}^C$ also affects. This includes MHD stability near the axis\cite{landreman2020,rodriguez2023mhd}, the shaping of flux surfaces\cite{rodriguez2023mhd}, and as a result, the smallest effective aspect ratio of the configuration\cite{landreman2021a}. Characterising the influence of second-order parameters on geometry in a clear way is non-trivial. Detailed analysis on shaping was presented as part of [\onlinecite{rodriguez2022mhd}], which we refer the reader to for a detailed discussion. However, one can show that in the large $|B_{22}^C|$ limit, flux surfaces become increasingly shaped to limit their achievable aspect ratio. For this to be lower than 10, we shall limit, quite crudely, $|B_{22}^C|\lesssim10$ (see Appendix~\ref{sec:appB2cBound}). This is only a rough estimate, but it provides a useful domain to perform the search of $B_{22}^C$ to minimise $\Delta B_{20}$. This will leave out any good QS configuration outside the allowed range of $B_{22}^C$. To include some of those cases, one may proceed a posteriori by relaxing the parameter choices.
\par

\section{Space of quasisymmetric configurations}
Following the arguments in the previous sections, we have a prescription to complete a second-order near-axis model for every regular axis shape. Every point in our space of curves represents a stellarator-symmetric, vacuum field stellarator that tries to be as quasisymmetric as possible while preserving some minimal requirements on the rotational transform and shaping. To illustrate the power of the approach, we consider the space of configurations spanned by magnetic axes described by two Fourier harmonics. We do not keep $\{Z_n\}$ explicitly in this space, and instead, for each set $\{R_n\}$ we look for the most quasisymmetric solution performing a Nelder-Mead optimisation [\onlinecite[Ch.~9.5]{wright1999}] on $\{R_n\}$ under the constraint that $Z_n\sim R_n$. This is consistent with the evidence gathered from optimised QS designs. This way, each point in the $\{R_n\}$ space corresponds to an optimal QS axis shape. 
\par
\begin{figure*}
    \centering
    \includegraphics[width=\textwidth]{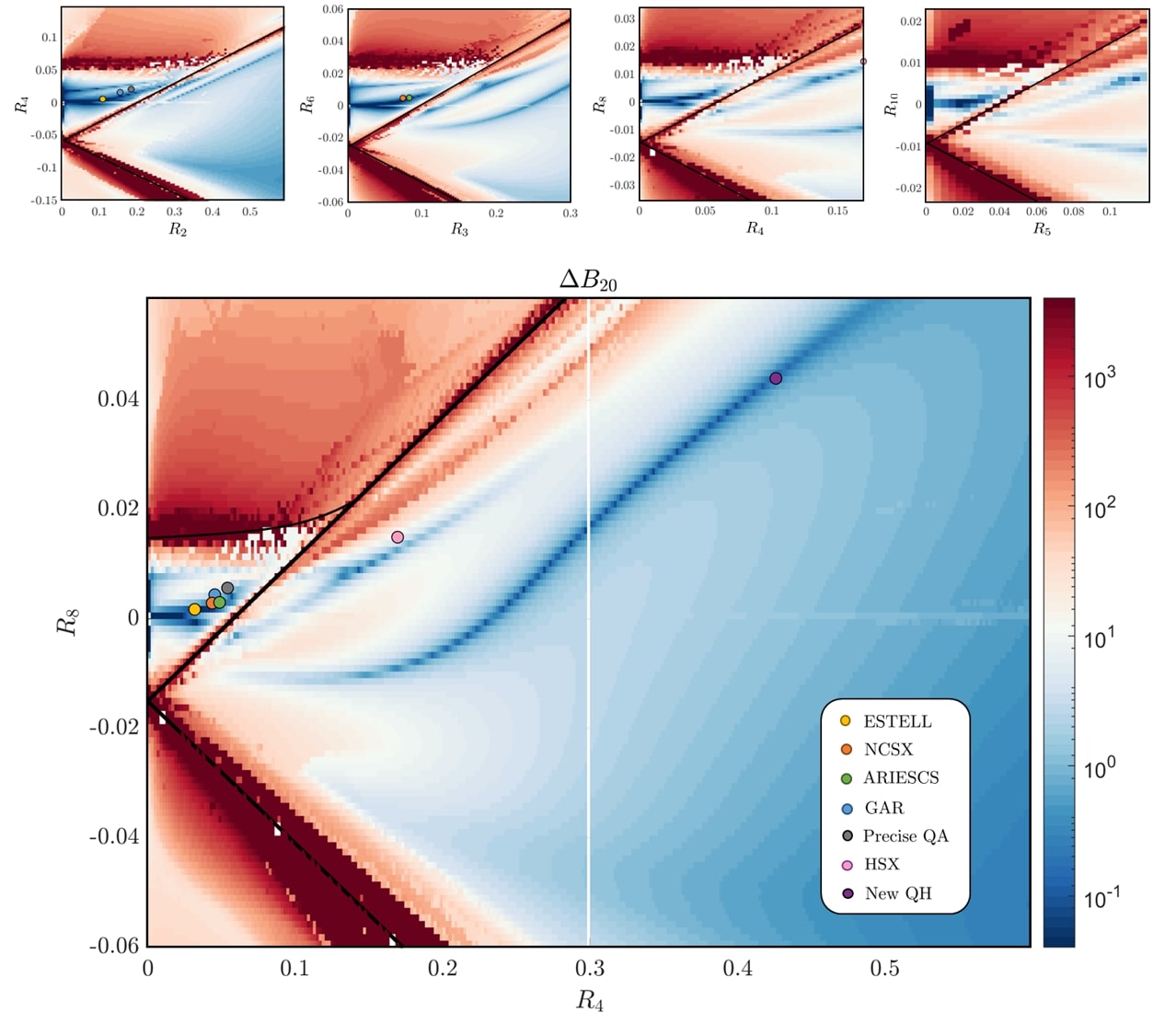}
    \caption{\textbf{Quality of quasisymmetry for the two-harmonic quasisymmetric configuration space.} The figure shows $\Delta B_{20}$ (in logarithmic scale) for the space of configurations spanned by two harmonic magnetic axes for several field periods $N=2,~3,~4,~5$ in the top figures and $N=4$ for the lower one. As the text describes, the $\{Z_n\}$ harmonics have been optimised at each point of space to minimise the quasisymmetry residual. The coloured scatter points represent typical quasisymmetric designs in our reduced space (see legend). To represent them all in the $N=4$ space, the magnetic axis harmonics are rescaled as $R_{4n}=R_{nN}(1+n^2N^2)/(1+16n^2)$ following the insight in [\onlinecite{rodriguez2022phase}]. Typical designs lie close to the bands of good quasisymmetry, which show the power of the approach. The black lines represent phase transition curves for $R_n=Z_n$. The dark purple point represents a new QH design, construction presented in Figure~\ref{fig:newQHexample}. The gap at $R_4$ is numerical, as the numerical evaluation of the space was performed in two separate runs.}
    \label{fig:spaceQSquality}
\end{figure*}

This reduced space is adequate to capture common QS designs, as was explicitly shown in [\onlinecite{rodriguez2022phase}], and may be seen in Figure~\ref{fig:spaceQSquality}. The spaces for a number of different field periods are presented with the colormap representing the quality of QS, $\Delta B_{20}$. The spaces were generated using the C++ libraries \texttt{qsc}\footnote{See https://github.com/landreman/qsc. The script used to obtain the main plot in Fig.~\ref{fig:spaceQSquality} can be found in the Zenodo repository associated to this paper. The same may be achieved, albeit slower, using \texttt{pyQSC}, which was how it was originally done and is also included there.} and \texttt{gsl} for the near-axis calculations and optimisation, respectively. On average, the evaluation of each point in this space takes less than a second running on a single CPU in a laptop\footnote{An 11th Gen i7-11850H core was used for this purpose. The main space in Fig.~\ref{fig:spaceQSquality} (which is 300x300) took a total of about 14~hrs. Most time is devoted to the optimisation sub-problems at each point (search for $\eta^*$, $B_{22}^C$, and $\{Z_n\}$). Of course, the construction of such a space is trivially parallelisable.}
\par
The space exhibits two remarkable features. First, the QS phase structure studied in the context of the magnetic axis makes itself clear (see phase transitions as solid black lines). Such features, as well as others, remain largely unchanged as the number of field periods changes. This allows us to represent the QS designs in Table~\ref{tab:compEtadesigns} together in Fig.~\ref{fig:spaceQSquality}. The second important feature of this configuration space is the appearance of what we call \textit{quasisymmetric branches}. These branches consist of well-distinguished regions of configuration space with excellent QS. (We leave a more precise definition for the future, a definition that will be necessary for a more systematic study of the branches.) Importantly, these branches agree with the location of typical QS designs.\cite{Najmabadi2008,Drevlak2013,Garabedian2008,Garabedian2009,Anderson1995,Zarnstorff2001,Ku2010,landreman2022}. It proves the predictive power of the approach and the role of our model as a unifying framework.
\par
With the QS branches identified, we have all the tools from the near-axis framework to investigate their properties. To illustrate what can be learned from such an analysis, let us focus on the dominant branch in the QA phase (the phase that includes the origin) that grows from the origin in the direction of the lowest harmonic $R_n$. Many (if not all) standard QA designs belong to this branch, irrespective of $N$.
\par
\begin{figure*}
    \centering
    \includegraphics[width=\textwidth]{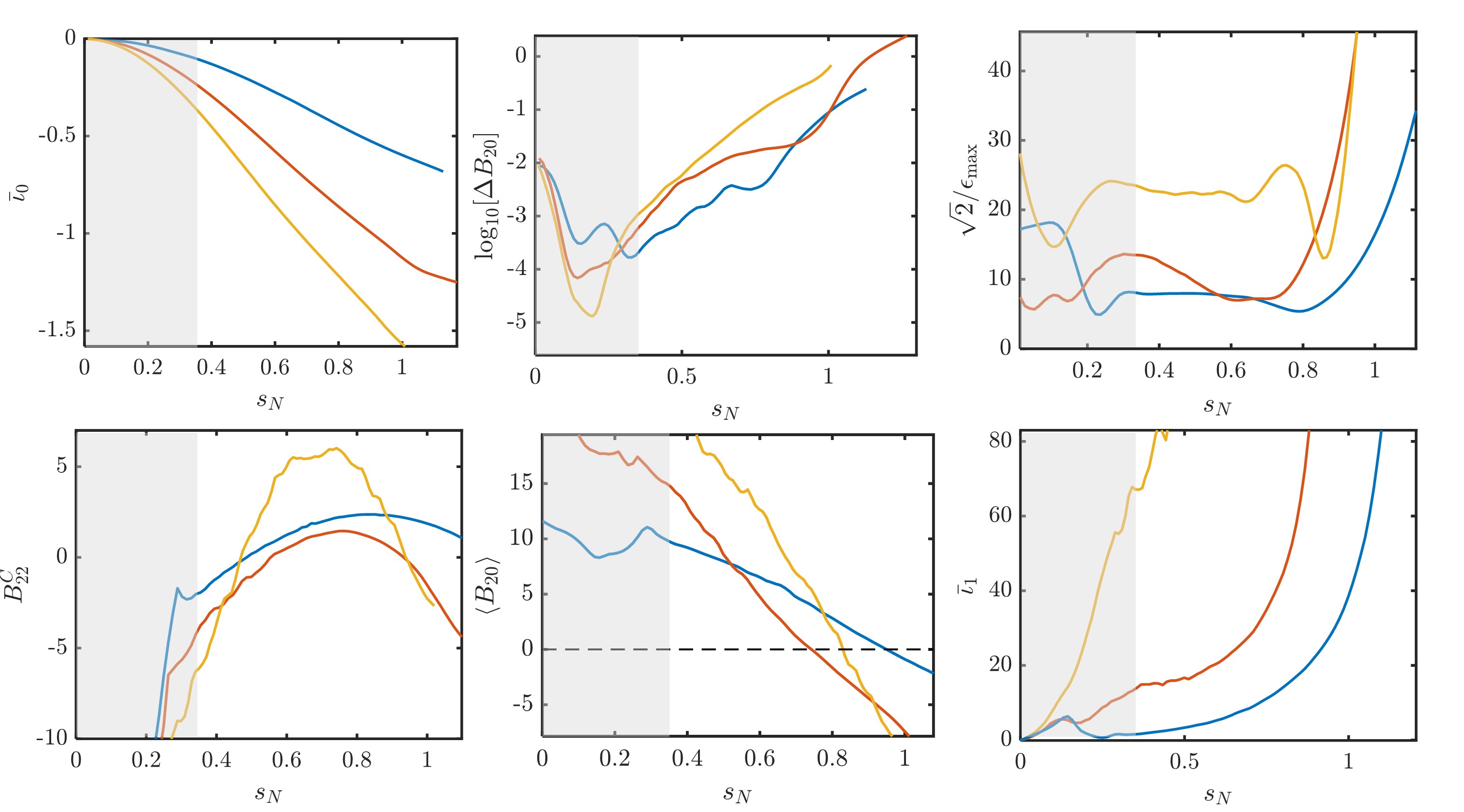}
    \caption{\textbf{Properties of the configurations in the main QA branch.} Properties of configurations in the main QA branch for $N=2,~3,~4$ (blue, orange, yellow respectively) plotted against\cite{rodriguez2022phase} $s_N=R_N(1+N^2)$. The properties shown are the rotational transform on the axis, the QS residual, the limit on aspect ratio, the $B_{22}^C$ parameter, the average $B_{20}$ for the magnetic well criterion, and the magnetic shear. The latter is defined so that the change in rotational transform is $\bar{\iota}_1$ times the inverse aspect ratio squared. These are obtained by following the branches in the 2-harmonic space of Fig.~\ref{fig:spaceQSquality}. The shaded area indicates a region where the construction of the NAE model is not well-behaved due to proximity to axisymmetry (see main text).}
    \label{fig:QAbranch}
\end{figure*}
Without delving into the origin of the branches, which we leave for future work, we now describe the properties and trends of this class of configurations as evident from Fig.~\ref{fig:QAbranch}. These properties should be interpreted as representative of the class, noting that many features may be changed by additional tweaking of the axis shape and parameter choice. However, before looking at these, note that in the region closest to the origin (i.e., near axisymmetry), the model presents ill behaviour, as it fails to show $\Delta B_{20}\rightarrow0$. Such misbehaviour aligns with the uncertainty in the choice of $B_{22}^C$ that arises close to axisymmetry (see $B_{22}^C$ in Fig.~\ref{fig:spaceQSquality}). The reason for this type of behavior may be deeper than we choose to pursue here, see [\onlinecite{plunk2018}].
\par
Away from this region, some physically interesting trends are observed. The QS quality degrades as the QA-QH phase transition is approached and the rotational transform and the magnetic shear grow. The degradation may be seen as a result of an increasingly shaped configuration driven by an increasingly twisted axis, a behaviour predicted in [\onlinecite{rodriguez2022phase}]. This also leads to an increase in the minimum effective aspect ratio $\sqrt{2}/\epsilon_\mathrm{max}$, as it does with the number of field periods. An aspect ratio of roughly $5-7$ appears possible for $N=2$ but increases to $15-25$ for $N=4$. Thus, although the rotational transform grows with $N$, which is of interest, the limitation in the compactness (and the quality of QS) restricts the configurations of interest to the lower $N$ values, as observed in optimisation efforts\cite{Zarnstorff2001,landreman2022}. Compared to the rest of the QA phase space, the shear, the QS residual, and $1/\epsilon_\mathrm{max}$ are small along the QA branch. This appears to match the observation that the magnetic shear of QS configurations tends to be small in practice\cite{landreman2022}. The magnetic shear also presents a sign opposite to that of $\iota_0$, and thus the rotational transform profiles are tokamak-like: the rotational transform decreases towards the edge of the configuration. The magnetic well criterion\cite{landreman2020,rodriguez2023mhd} (via the sign of $\langle B_{20}\rangle$) shows that the majority of the QA branch gives rise to a magnetic hill, and thus is MHD unstable. This aligns with the conclusions reached in [\onlinecite{landreman2022}], where an additional effort was made to reach a solution with a magnetic well. Those configurations along the QA branch further away from the origin will be more easily stabilised by additional tweaking of the configuration since their magnetic hill is shallow. 
\par
This discussion of the QA branch is only a partial account of the full power of the present approach to understand optimised QA configurations. A similar effort could be devoted to the branches in the QH phase, but we shall not do that here. Instead, we content ourselves with a few general observations. We call the branch the \textit{HSX branch} as it lies close to the HSX design\cite{Anderson1995}. The other branch appears not to include any existing QH design, and thus we shall refer to it as the \textit{new QH branch}. The QH space is significantly more sensitive to the parameter choices than the QA phase, which in particular leads to the location of the new QH branch in phase space changing under different choices of $\eta$. If $\eta$ were to be treated as another optimisation parameter, the branch would become a broader region bounded by large elongation configurations. This latter branch is particularly interesting. Configurations along it lack the common bean-shaped cross-section and exhibit a natural magnetic well. Nevertheless, most remarkably, no existing QS-optimised design belongs to this class. This illustrates the power of this approach in exploring QS-optimised configurations. 
\par
\begin{figure*}
    \centering
    \includegraphics[width=\textwidth]{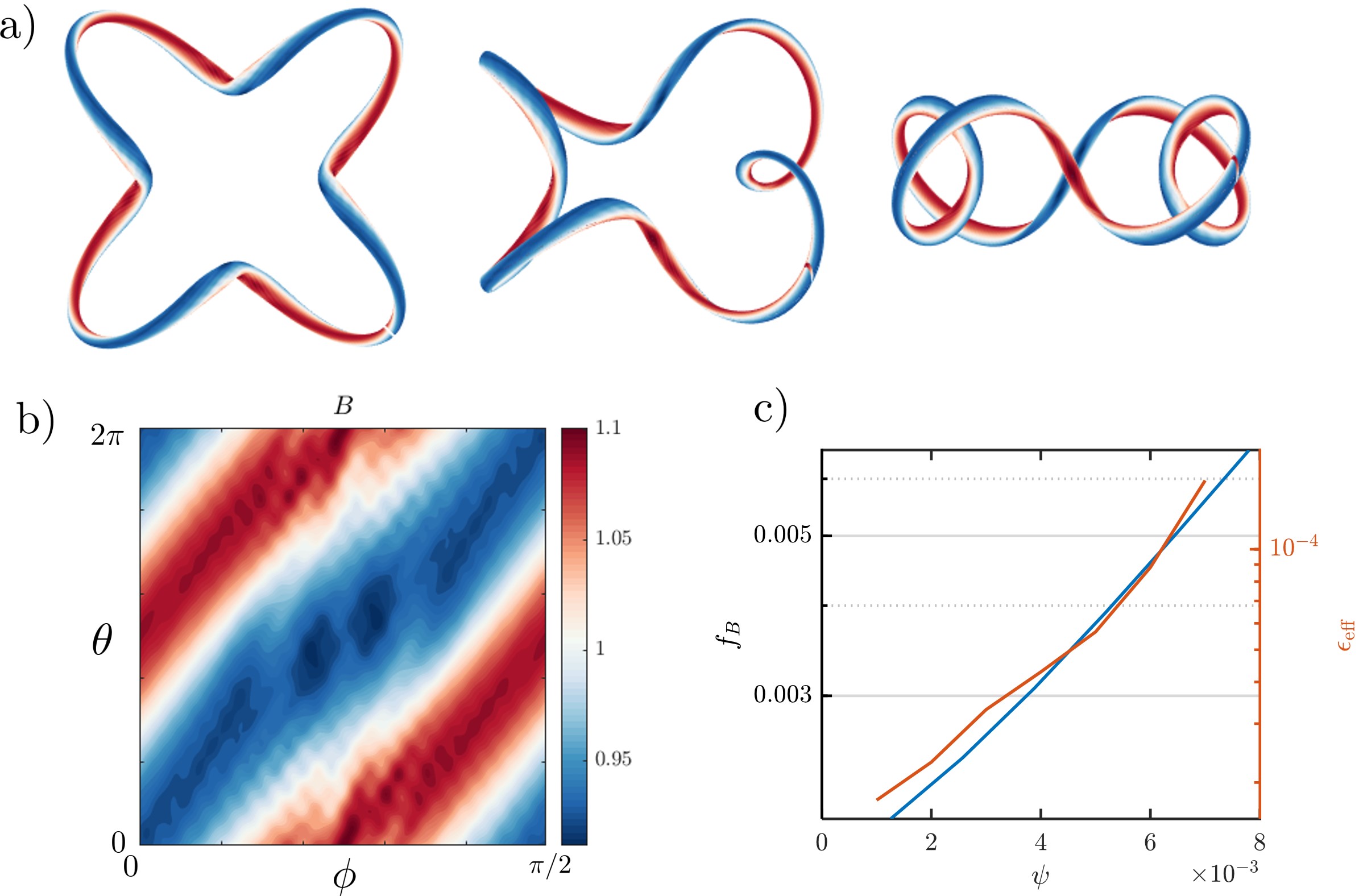}
    \caption{\textbf{Example of new QH branch configuration.} a) Different projections of the 3D boundary of a configuration from the new QH branch. The colormap represents contours of constant $|\mathbf{B}|$. The global equilibria were solved using \texttt{VMEC} given the surface of an NAE construction with an axis given by the following $\{R_n,Z_n\}$ components $R_n=\{0.426,~0.044,~-6.36\times 10^{-11},~2.85\times 10^{-5},~3.89\times 10^{-8}\}$ and $Z_n=\{0.411,~0.043,~6.53\times 10^{-5},~1.36\times 10^{-5},~1.16\times 10^{-5}\}$. The higher harmonics have been chosen as small last tweaking choices to minimise further $\Delta B_{20}$, which within the QS framework, we make $\Delta B_{20}\sim7\times 10^{-3}$. b) Magnetic field magnitude on the last flux surface of the global solution. c) Quasisymmetric residual and $\epsilon_\mathrm{eff}$ (a measure of particle transport) \cite{nemov1999,nemov2014} as a function of radius in the global solution showing the quasisymmetric nature of the configuration.}
    \label{fig:newQHexample}
\end{figure*}

We present for completeness an example of a stellarator belonging to this branch, which we call `new QH'. We construct a global equilibrium solving the equilibrium problem inside a fixed outer surface, constructed from a finite aspect ratio evaluation of the near-axis model (see Fig.~\ref{fig:spaceQSquality}) following [\onlinecite{landreman2019}]. This form of linking the near-axis and global solutions is not the best, as it uses the worst-described feature of the near-axis model (the `outer' surface) as an intermediary. However, it suffices as a first approximation. The particular configuration in the space of Fig.~\ref{fig:spaceQSquality} was chosen to allow for a reasonable aspect ratio (an aspect ratio of $A\sim13.5$ given by \texttt{VMEC}, which has difficulty initialising the solver in more compact scenarios). To construct the final form of the configuration, some additional refinement of the axis shape was made within the near-axis framework by allowing for three small additional harmonic components. The configuration in Fig.~\ref{fig:newQHexample} shows good QS behaviour as expected from the near-axis model. This is especially important, considering that no optimisation has been performed in the space of global equilibria. 
\par
This configuration serves as an example of the approach's potential. The complications encountered by the numerical solvers (including the large axis excursion) could be a reason behind the design efforts missing out stellarators belonging to the new QH branch. However, the exact reason is hard to pinpoint. It could have also resulted from other constraints (such as aspect ratio) or the initial guesses 
\par
Other authors have recently found stellarator designs considering optimisation in the space of near-axis configurations\cite{landreman2022map}. The result is a multitude of designs, many of which have previously unseen features. The approach there may be regarded complementary to that presented here, as this should serve to structure, guide (e.g., seeding the search to be more exhaustive), and understand the other approach. 

\section{Conclusion}
In this paper, we have presented the construction of a model for quasisymmetric stellarators based on the near-axis expansion, in which configurations can be identified with their axis shapes only. Doing so enables us to represent the space of configurations in a form that inherits the topological structure of a space of closed regular curves. To concoct such a model requires a careful choice of parameters that form part of the near-axis expansion.
\par
In the case of vacuum field, stellarator-symmetric stellarators (although the extension to more general cases should be straightforward), there are two such parameters: $\eta$ and $B_{22}^C$. The former is chosen to guarantee that flux surfaces are not extremely elongated, maximising the rotational transform in quasiaxisymmetric configurations. The choice as presented always exists and is unique. The $B_{22}^C$ parameter is then chosen so that each axis shape is represented by the `most quasisymmetric' configuration. That way, we can construct an example space of quasisymmetric stellarators. 
\par
The model constructed is a powerful tool that reproduces stellarator designs optimised for quasisymmetry by other approaches. It does so naturally, grouping them into families that we call quasisymmetric branches. We show explicitly how the model may be used to study the properties of one such branch, namely the main QA branch. We also showcase how new configurations appear from this approach. The structure of the space and simplicity of the model opens the door to a fundamental study of quasisymmetric stellarators and their properties, as well as a practical exploration of designs complementary to recent attempts\cite{landreman2022map}. However, there remains significant room for future work, including a systematic way of identifying branches, initialising optimisation from these, and a more robust way of connecting the near-axis construction to global equilibrium solvers. 

\hfill

\section*{Acknowledgements}
The authors would like to acknowledge fruitful discussions with M. Landreman, R. Jorge, R. Nies, S. Buller, and E. Paul.
This research was primarily supported by a grant from the Simons Foundation/SFARI (560651, AB) and DoE Contract No DE-AC02-09CH11466. ER was also partially supported by the Charlotte Elizabeth Procter Fellowship at Princeton University. 

\section*{Data availability}
The data supporting this study's findings are available at the following Zenodo repository \texttt{https://doi.org/10.5281/zenodo.7817884}.

\appendix

\section{Definition and uniqueness of $\eta^*$} \label{sec:appEtaStar}
The choice of $\eta$ modifies the competition between the different contributions to the rotational transform on the axis. We learned in the main text that at large and small $\eta$, the rotational transform $\bar{\iota}_0$ vanishes. Thus, by the mean value theorem under the assumption of continuity, it must be that the function $\bar{\iota}_0(\eta)$ has at least a turning point. Let us see what we can learn about this point.
\par
Consider linearising the Riccati equation, Eq.~(\ref{eqn:sigmaEqn}), with respect to $\eta$, 
\begin{widetext}
\begin{equation}
    \frac{\mathrm{d}}{\mathrm{d}\phi}\frac{\delta\sigma}{\delta\eta}=-2\bar{\iota}_0\sigma\frac{\delta\sigma}{\delta\eta}-\bar{\iota}_0\frac{\eta^3}{B_0\kappa^4}+\frac{\eta B_{\alpha0}}{\kappa^2}(2\tau+B_{\theta20})-\frac{\delta\bar{\iota}_0}{\delta\eta}\left[1+\sigma^2+\frac{1}{4B_0}\left(\frac{\eta}{\kappa}\right)^4\right]. \label{eqn:firstVariationRiccati}
\end{equation}
\end{widetext}
Looking at the extremum $\delta\bar{\iota}_0/\delta\eta\stackrel{!}{=}0$ makes Eq.~(\ref{eqn:firstVariationRiccati}) an ODE on $\delta\sigma/\delta\eta$, the change in the solution $\sigma$ upon infinitesimal change of the parameter $\eta$. Thus, $\delta\sigma/\delta\eta$ must, as $\sigma$, be periodic for $\sigma+\delta\sigma$ to be so. Because $\sigma(0)=0$ is kept fixed, $\delta\sigma/\delta\eta=0$ at $\phi=0$. The linearised ODE can then be solved by an integration factor,
\begin{widetext}
\begin{equation}
    \frac{\delta\sigma}{\delta\eta}=e^{-2\bar{\iota}_0\int_0^\phi\sigma\mathrm{d}\phi'}\int_0^\phi e^{2\bar{\iota}_0\int_0^{\phi'}\sigma\mathrm{d}\phi''}\frac{2}{\eta}\left[\frac{B_{\alpha0}}{2}(2\tau+B_{\theta20})\left(\frac{\eta}{\kappa}\right)^2-\frac{\bar{\iota}_0}{2B_0}\left(\frac{\eta}{\kappa}\right)^4\right]\mathrm{d}\phi'. \label{eqn:varSigma}
\end{equation}  
\end{widetext}
Imposing periodicity, using the Riccati equation, Eq.~(\ref{eqn:sigmaEqn}), and integrating by parts (assuming $\bar{\iota}_0\neq0$ and stellarator symmetry to drop the boundary term), we obtain 
\begin{equation}
    \int_0^{2\pi}\left[\sigma^2+\frac{1}{4B_0}\left(\frac{\eta}{\kappa}\right)^4-1\right]E\mathrm{d}\phi=0, \tag{\ref{eqn:extrEtaIotaRel}}
\end{equation}
where $E=\exp[{2\bar{\iota}_0\int_0^{\phi}\sigma\mathrm{d}\phi'}]$. 
\par
\hrulefill\par
\textbf{Definition of $\eta^*$:}
    the first-order parameter $\eta$ extremises the rotational transform on the axis of a stellarator symmetric construction iff,
    \begin{equation}
        \int_0^{2\pi}\left[\sigma^2+\frac{1}{4B_0}\left(\frac{\eta}{\kappa}\right)^4-1\right]E\mathrm{d}\phi=0, \tag{\ref{eqn:extrEtaIotaRel}}
    \end{equation}
    where $E=\exp[{2\bar{\iota}_0\int_0^{\phi}\sigma\mathrm{d}\phi'}]$.
    \par
\hrulefill\par
Although exact, the condition in Eq.~(\ref{eqn:extrEtaIotaRel}) is implicit through $\sigma$. This prevents a closed form of $\eta^*$, which has to be found numerically. To this end, finding bounds on $\eta^*$ from Eq.~(\ref{eqn:extrEtaIotaRel}) is helpful. Because $\sigma^2$ and $E$ are both positive quantities, in order for Eq.~(\ref{eqn:extrEtaIotaRel}) to hold, the integrand in square brackets must cross zero somewhere. However, when $\eta\geq(4B_0)^{1/4}\kappa_\mathrm{max}$, the integrand is always positive, and thus there cannot be a solution. This serves as an upper bound on $\eta^*$. In the small $\eta$ limit (with $\sigma\sim\eta^2$), the $-1$ piece dominates, and the equation cannot be satisfied. We may then rigorously give the interval $\eta^*\in(0,\sqrt{2}\kappa_\mathrm{max}B_0^{1/4}]$, or with less rigour, change the lower bound to $\eta^*>\sqrt{2}\kappa_\mathrm{min} B_0^{1/4}$, condition below which $\sigma^2$ in Eq.~(\ref{eqn:extrEtaIotaRel}) is the only term that may balance the other two. Given the differences in curvature between the QA and QH phases (see [\onlinecite{rodriguez2022phase}]), $\eta^*$ will tend to be larger in QH configurations. 
\par
Although the existence of $\eta^*$ is guaranteed by the form of the $\eta$ asymptotes and the mean value theorem, we do not know whether such an extremum is unique. To prove so, we will investigate the second variation of the Riccati equation, Eq.~(\ref{eqn:sigmaEqn}), and its sign at $\eta=\eta^*$. Taking the variation of Eq.~(\ref{eqn:firstVariationRiccati}) (and looking at the extrema),
\begin{multline}
    \frac{\mathrm{d}}{\mathrm{d}\phi}\frac{\delta^2\sigma}{\delta\eta^2}=-2\bar{\iota}_0\left[\left(\frac{\delta\sigma}{\delta\eta}\right)^2+\sigma\frac{\delta^2\sigma}{\delta\eta^2}\right]-3\bar{\iota}_0\frac{\eta^2}{B_0\kappa^4}+\\
    +\frac{B_{\alpha0}}{\kappa^2}(2\tau+B_{\theta20})-\frac{\delta^2\bar{\iota}_0}{\delta\eta^2}\left[1+\sigma^2+\frac{1}{4B_0}\left(\frac{\eta}{\kappa}\right)^4\right]. \label{eqn:secondVariationRiccati}
\end{multline}
The periodicity requirement in $\delta^2\sigma/\delta\eta^2$ may be written in the  following solvability form: eliminating $\tau$ using the Riccati $\sigma$-equation, Eq.~(\ref{eqn:sigmaEqn}), integrating by parts, using the $\eta^*$ condition, Eq.~(\ref{eqn:extrEtaIotaRel}), and using the odd parity of $\sigma$ in stellarator symmetry, we obtain 
\begin{equation}
    \int_0^{2\pi} E\left\{-2\bar{\iota}_0\left[\left(\frac{\delta\sigma}{\delta\eta}\right)^2+\frac{\eta^2}{B_0\kappa^4}\right]-P\frac{\delta^2\bar{\iota}_0}{\delta\eta^2}\right\}\mathrm{d}\phi=0,
\end{equation}
where $P=1+\sigma^2+(\eta/\kappa)^4/4B_0$. The term with the square bracket has a sign of $-\mathrm{sgn}(\bar{\iota}_0)$. As $P, E>0$, for the integral to vanish, it must be the case that the sign of the second variation of $\bar{\iota}_0$ (which is not a function of $\phi$) satisfies,
\begin{equation}
    \mathrm{sgn}\left(\frac{\delta^2\bar{\iota}_0}{\delta\eta^2}\right)=-\mathrm{sgn}(\bar{\iota}_0).
\end{equation}
As the sign of $\bar{\iota}_0$ is set by the sign of the combination $\int (B_{\alpha0}/\kappa^2)(2\tau+B_{\theta20})\mathrm{d}\phi$ (and thus cannot change with $\eta$), the extrema of $\bar{\iota}_0$ can only be either maxima or minima (but only one of these), as the sign of $\delta^2\bar{\iota}_0/\delta\eta^2$ is fixed. Thus, $\eta^*$ is unique. 

\section{Magnetic shear decoupling for $\eta^*$} \label{sec:appShear}
It was shown in [\onlinecite{rodriguez2021weak}] that, within the near-axis framework, one could evaluate magnetic shear (and higher derivatives of rotational transform) as solvability conditions of first-order periodic ODEs, so-called \textit{generalised $\sigma$ equations}. Physically, this is reasonable since the rotational transform must be self-consistently chosen given an average toroidal current profile. Here we shall focus on magnetic shear, writing $\bar{\iota}=\bar{\iota}_0+\psi\bar{\iota}_1+\dots$, $\mathrm{d}\bar{\iota}/\mathrm{d}\psi=\bar{\iota}_1$ [\onlinecite[Sec.~2.8]{Helander2014}]. Having an understanding of this quantity is important, as it affects important properties of the stellarator, such as ballooning stability [\onlinecite[Ch.~6.14]{wessonTok}]. 
\par
From [\onlinecite{rodriguez2021weak}], it is clear that magnetic shear is a third-order quantity in the near-axis framework (or rather 2.5 order) in a vacuum given by 
\begin{multline}
    \bar{\iota}_1=\frac{\int_0^{2\pi}\mathrm{d}\phi' e^{2\bar{\iota}_0\int_0^{\phi'}\sigma\mathrm{d}\phi''}\tilde{\Lambda}_3}{\int_0^{2\pi}\mathrm{d}\phi'e^{2\bar{\iota}_0\int_0^{\phi'}\sigma\mathrm{d}\phi''}\left[1+\sigma^2+\frac{1}{4B_0}\left(\frac{\eta}{\kappa}\right)^4\right]}. \label{eqn:condIotaGenSigma}
\end{multline}
where,
\begin{align}
    \tilde{\Lambda}_3=&\frac{1}{(Y^S_{1,1}){}^2} \left\{2 l'\left[\tau \left(X^C_{3,1}Y^S_{1,1}+2 X^C_{2,2}Y^S_{2,2}+X^C_{1,1}Y^S_{3,1}-\right.\right.\right.\nonumber\\
    &\left.\left.\left.X^S_{3,1}Y^C_{1,1}-2 X^S_{2,2}Y^C_{2,2}\right)+\kappa \left(2 X^C_{2,2}Z^S_{2,2}+X^C_{1,1}Z^S_{3,1}- \right.\right.\right.\nonumber\\
    &\left.\left.\left.-2 X^S_{2,2}Z^C_{2,2}\right)\right]-2 \iota _0 \left(2 X^C_{2,2}{}^2+X^C_{1,1}X^C_{3,1}+2 X^S_{2,2}{}^2+\right.\right.\nonumber\\
    &\left.\left.+2 Y^C_{2,2}{}^2+2 Y^S_{2,2}{}^2+Y^S_{1,1}Y^S_{3,1}+2 Z^C_{2,2}{}^2+2 Z^S_{2,2}{}^2\right)-\right.\nonumber\\
    &\left.-X^S_{3,1}X^C_{1,1}{}'-2 X^S_{2,2}X^C_{2,2}{}'+2 X^C_{2,2}X^S_{2,2}{}'+X^C_{1,1}X^S_{3,1}{}'\right.\nonumber\\
    &\left.-Y^S_{3,1}Y^C_{1,1}{}'-2 Y^S_{2,2}Y^C_{2,2}{}'+2 Y^C_{2,2}Y^S_{2,2}{}'+Y^C_{1,1}Y^S_{3,1}{}'\right.\nonumber\\
    &\left.-2 Z^S_{2,2}Z^C_{2,2}{}'+2 Z^C_{2,2}Z^S_{2,2}{}'\right\}. \label{eqn:Lam3}
\end{align}
The weighted integral of this quantity $\tilde{\Lambda}_3$ drives the magnetic shear, Eq.~(\ref{eqn:condIotaGenSigma}). 
\par 
This quantity depends on third-order quantities, for which closed forms may be found in the literature\cite{rodriguez2020i,landreman2019,garrenboozer1991b}. Although Eq.~(\ref{eqn:Lam3}) is of the third order, most elements in it may be written exclusively in terms of lower order quantities, except for $X_3$, which explicitly introduces parameters $B_{31}^C$ and $B_{31}^S$. Therefore, let us focus on the elements in the shear that depend on $B_{31}^C$. 
\par
This component of the magnetic field may be seen as the variation of the parameter $\eta$ with radius. An approximate variation of ellipticity of flux surfaces with radius. It is then unsurprising that, given the central role of $\eta$ in determining the rotational transform on the axis, $B_{31}^C$ directly affects the shear. Explicitly, the dependence of $\tilde{\Lambda}_3$ on $B_{31}^C$ is,
\begin{equation}
    \tilde{\Lambda}_3=\left[\frac{\bar{\iota}_0}{B_0\eta}\left(\frac{1}{4B_0}\left(\frac{\eta}{\kappa}\right)^4-1\right)-\frac{\sigma'}{2B_0\eta}\right]B_{31}^C+\dots, \label{eqn:Lam3B31}
\end{equation} 
from the expressions for $X_{31}$ (see Appendix F in [\onlinecite{rodriguez2021weak}]). Define the factor $\mathcal{M}$ as the multiplicative factor modulating the contribution of $B_{31}^C$ to the magnetic shear. Using Eq.~(\ref{eqn:Lam3B31}) in Eq.~(\ref{eqn:condIotaGenSigma}), integrating by parts and assuming stellarator symmetry, we obtain
\begin{equation}
    \mathcal{M}=\frac{\bar{\iota}_0}{B_0\eta}\left[1-\frac{2\int_0^{2\pi} e^{2\bar{\iota}_0\int\sigma\mathrm{d}\phi'}\mathrm{d}\phi}{\int_0^{2\pi}\left[1+\sigma^2+\frac{1}{4B_0}\left(\frac{\eta}{\kappa}\right)^4\right] e^{2\bar{\iota}_0\int\sigma\mathrm{d}\phi'}\mathrm{d}\phi}\right]. \label{eqn:MfactorShear}
\end{equation}
The expression in the large square brackets is a number between $(-1,1)$ sets a hard upper bound to the effect of the third-order modulation on the shear. Furthermore, it restricts the rotational transform on the axis not to exceed $(B_{31}^C/B_0)/\eta$. The minimum of $\mathcal{M}$ with finite (non-zero) $\eta$ is achieved when the expression in brackets vanishes. Rewriting it into a single fraction, we get
\begin{equation}
    \mathcal{M}=\frac{\bar{\iota}_0}{B_0\eta}\frac{\int_0^{2\pi}\left[\sigma^2+\frac{1}{4B_0}\left(\frac{\eta}{\kappa}\right)^4-1\right] E\mathrm{d}\phi}{\int_0^{2\pi}\left[1+\sigma^2+\frac{1}{4B_0}\left(\frac{\eta}{\kappa}\right)^4\right] E\mathrm{d}\phi}, \label{eqn:MfactEta}
\end{equation}
where $E=\exp[2\bar{\iota}_0\int\sigma\mathrm{d}\phi']$. The numerator is precisely the extremum condition for $\eta^*$ in Eq.~(\ref{eqn:extrEtaIotaRel}). Thus, the resilience to $\eta$ presents itself by making the magnetic shear independent of third-order choices. Thus, $\iota_1$ becomes a second-order quantity upon choosing $\eta^*$. This makes the truncated near-axis model, as constructed in this paper, more complete. Note that $\mathcal{M}$ vanishing does not equal vanishing of the magnetic shear. There remains a generally non-zero `intrinsic' contribution from lower-order pieces. However, if configurations with a reasonably low aspect ratio are sought, then the second-order shaping should remain small, and so will the magnetic shear. 

\section{Choosing $\eta$ to maximise $L_\nabla$} \label{sec:appGradB}
Let us consider briefly in this Appendix some of the intricacies of the measure proposed in [\onlinecite{landreman2021a}] as a guiding principle in the choice of the parameter $\eta$. The measure is the magnitude of $||\nabla\mathbf{B}||$, using the Frobenius norm $||M_{ij}||^2=\sum_{i,j}(M_{ij})^2$ of the $\nabla\n{B}$ tensor. This provides a characteristic length scale of the field, approximately that corresponding to the maximum distance at which a coil may be placed\cite{landreman2021a}. 
This measure has been used as a proxy of the radius of applicability of the near-axis construction, which appears to work when applied to optimisation within the near-axis framework in practice\cite{landreman2022map,jorge2022}. 
\par
From this perspective, it appears to be a natural choice of $\eta$ that maximises the length scale $L_\nabla\stackrel{\cdot}{=}1/||\nabla\n{B}||$. To investigate this, we write the gradient of $\n{B}$ at first order (in the notation used in this paper),
\begin{equation}
    \nabla\mathbf{B}=\nabla\psi\partial_\psi\mathbf{B}+\nabla\chi\partial_\chi\mathbf{B}+\nabla\phi\partial_\phi\mathbf{B}.
\end{equation} 
Using $\mathbf{B}=\mathcal{J}^{-1}(\partial_\phi+\bar{\iota}\partial_\chi)\mathbf{x}$ and the dual relations\cite{garrenboozer1991a}, we may then expand the expression in the near-axis fashion,
\begin{multline}
    \nabla\mathbf{B}\approx\mathcal{J}_0^{-1}\partial_\chi\mathbf{x}_1\times\partial_\phi\mathbf{x}_0\partial_\psi\mathbf{B}_1+\mathcal{J}_0^{-1}\partial_\phi\mathbf{x}_0\times\partial_\psi\mathbf{x}_1\partial_\chi\mathbf{B}_1+\\
    \mathcal{J}_0^{-1}\partial_\psi\mathbf{x}_1\times\partial_\chi\mathbf{x}_1\partial_\phi\mathbf{B}_0.
\end{multline}
For $\mathbf{x}$ we take Eq.~(\ref{eqn:xDef}), to construct expressions like,
\begin{equation*}
    \partial_\chi\mathbf{x}_1\times\partial_\phi\mathbf{x}_0=\frac{\mathrm{d}l}{\mathrm{d}\phi}(\partial_\chi Y_1\hat{\kappa}-\partial_\chi X_1\hat{\tau}),
    \end{equation*}
    and
    \begin{multline*}
    \partial_\chi \mathbf{B}_1=\mathcal{J}^{-1}\left[-\kappa l'\partial_\chi X_1\hat{b}+(\partial_\chi Y_1'-\bar{\iota}Y_1-\tau l'\partial_\chi X_1)\hat{\tau}+\right.\\
    \left.(\partial_\chi X_1'-\bar{\iota}X_1+\tau l'\partial_\chi Y_1)\right].
\end{multline*}
With this,
\begin{multline}
    \nabla \mathbf{B}=\frac{1}{B_{\alpha0}B_0}\left[\frac{1}{2\sqrt{B_0}}\left(Y_{11}^SX_{11}^C{}'+\bar{\iota}_0Y_{11}^CX_{11}^C\right)\hat{\kappa}\hat{\kappa}+\right.\\
    \left.+\left(l'\tau-\frac{\bar{\iota}_0}{2l'}(X_{11}^C)^2\right)\hat{\tau}\hat{\kappa}+\right.\\
    \left.+\left(\frac{Y_{11}^C{}'Y_{11}^S-Y_{11}^S{}'Y_{11}^C+\bar{\iota}_0[(Y_{11}^S)^2+(Y_{11}^C)^2]}{2\sqrt{B_0}}-l'\tau\right)\hat{\kappa}\hat{\tau}+\right.\\
    +\left.\frac{1}{2\sqrt{B_0}}(X_{11}^CY_{11}^S{}'-\bar{\iota}_0X_{11}^CY_{11}^C)\hat{\tau}\hat{\tau}\right]+\frac{\kappa}{\sqrt{B_0}}(\hat{b}\hat{\kappa}+\hat{\kappa}\hat{b}). \label{eqn:nablaB}
\end{multline}
which is consistent with [\onlinecite[Eq.~(3.12)]{landreman2021a}], differences in notation provided. 
\par
Eq.~(\ref{eqn:nablaB}) may be interpreted as a function of $\eta$ for a fixed axis shape, much in the same way as we did for $\bar{\iota}_0$. As in that case, a basic understanding of the behaviour of $||\nabla\n{B}||$ can be gained from the large and small $\eta$ asymptotics. For stellarator symmetry and $\eta\rightarrow0$, (taking for ease of notation $B_0\sim1$),
\begin{equation}
    ||\nabla\mathbf{B}||=\sqrt{2\kappa^2+\frac{2}{(l')^2}\left(\frac{\kappa'}{\kappa}\right)^2+\tau^2+(\tau+B_{\theta20})^2}, \label{eqn:lowEtaGradB}
\end{equation}
while for large $\eta$,
\begin{widetext}
\begin{equation}
    ||\nabla\mathbf{B}||=\sqrt{2\kappa^2+\frac{2}{(l')^2}\left(\frac{\kappa'}{\kappa}-\sigma\bar{\iota}_0\right)^2+\left(\tau-\frac{\bar{\iota}_0}{2l'}\left(\frac{\eta}{\kappa}\right)^2\right)^2+\left(\tau-\frac{\bar{\iota}_0}{2l'}\left(\frac{\eta}{\kappa}\right)^2+B_{\theta20}\right)^2}. \label{eqn:largeEtaGradB}
\end{equation}
\end{widetext}
These are both generally different constant values, provided $\sigma\neq0$, which does not guarantee the existence of an extremum. An example of the special case is that of axisymmetry, for which the asymptotes are $||\nabla\mathbf{B}||\sim\sqrt{2+B_{\theta20}^2}$, and the extremum of $||\nabla\mathbf{B}||$ coincides with $\eta^*$, the circular cross-section. 
\par
\begin{figure*}
    \centering
    \includegraphics[width=\textwidth]{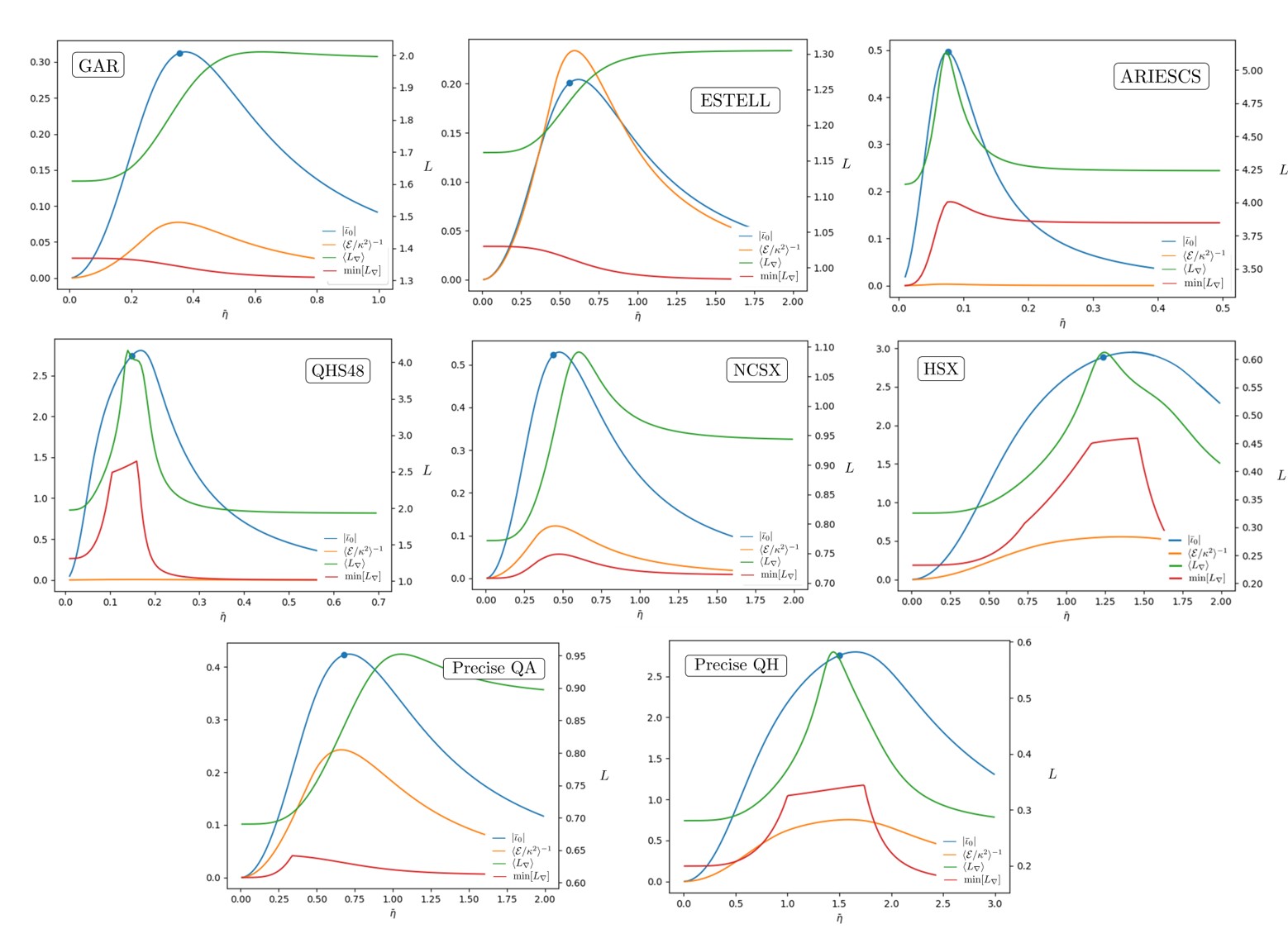}
    \caption{\textbf{Comparison of $\eta$ measures.} The main measures guiding the choice of $\eta$ discussed in the text are compared for the same configurations as Table~\ref{tab:compEtadesigns}. These include $|\bar{\iota}_0|$ whose extremum is $\eta^*$ (blue), the weighted elongation whose extremum is $\eta_{\bar{\mathcal{E}}}$ (orange), the mean $L_\nabla$ whose maximum is $\eta_{\nabla\n{B}}$ (green), and for comparison the minimum of $L_\nabla$ (red). Green and red correspond to the right-axis scale. The blue point represents $\eta$ from the global optimised designs, namely $\eta_\mathrm{VMEC}$. In many designs, all considerations provide similar $\eta$ values (see Table~\ref{tab:compEtadesigns}), but the shortcoming of $L_\nabla$ is also patent (especially in QA examples such as ESTELL \cite{Drevlak2013}). These also show the similarity in behaviour of $\langle\mathcal{E}/\kappa^2\rangle^{-1}$ and $|\bar{\iota}_0|$.}
    \label{fig:gradBmeasureothers}
\end{figure*}

The consequence of this uneven asymptotic behaviour is generally a small prominence of the extrema, if there is one. Examples of this behaviour are shown in Figure~\ref{fig:gradBmeasureothers}, where the $L_{\nabla}$ metric is compared to other first-order measures such as the rotational transform and the weighted averaged elongation for a selection of quasisymmetric designs. This uncertain behaviour of $||\nabla\n{B}||$, which shows excellent behaviour in the QH cases, makes this gradient length not as robust a measure as the other choices proposed (see Table~\ref{tab:compEtadesigns}). This excellent behaviour in the case of QH configurations may result from the tendency of such configurations to develop small-scale features. Like $\mathcal{E}$, $L_\nabla$ is also a function of $\phi$, and differences arise from different considerations. 

\section{Second order ODE on $B_{20}$} \label{sec:B20appendix}
The seed of overdetermination at the second order comes from the magnetic field having to satisfy both the QS and equilibrium conditions. The common way to avoid this overdetermination at the second order is to relax the condition of quasisymmetry partially. Doing so requires solving a second-order, regular ODE for $B_{20}$, the second-order, $\theta$ independent change in $1/|\mathbf{B}|^2$. We write this equation explicitly here for a stellarator-symmetric vacuum field:
\begin{equation}
    \mathcal{A}\frac{\mathrm{d}^2}{\mathrm{d}\phi^2}\left(\frac{B_{20}}{B_0}\right)+\mathcal{B}\frac{\mathrm{d}}{\mathrm{d}\phi}\left(\frac{B_{20}}{B_0}\right)+\mathcal{C}\frac{B_{20}}{B_0}+\mathcal{D}=0,
\end{equation}
where,
\begin{subequations}
\begin{gather}
    \mathcal{A}=-\frac{B_{\alpha0}\eta^2}{2\kappa^2\bar{\iota}_0l'}\left[1+\frac{4B_0\kappa^4}{\eta^4}(1+\sigma^2)\right], \\
    \mathcal{B}=\frac{2B_{\alpha0}\eta^2}{\bar{\iota}_0 l'}\frac{\kappa'}{\kappa^3}-\frac{4l'\sigma}{\bar{\iota}_0}\tau, \\
    \mathcal{C}=-\frac{l'}{2B_{\alpha0}\eta^2\kappa^2}\left[\bar{\iota}_0\left(4\kappa^4(1+\sigma^2)-\frac{3\eta^4}{B_0}\right)+8B_{\alpha0}\eta^2\kappa^2\tau\right],
\end{gather}
\end{subequations}
and $B_{\alpha0}^2B_0=(\mathrm{d}l/\mathrm{d}\phi)^2$. The homogeneous operator only depends on zeroth and first-order considerations. However, the inhomogeneous term $D$ is given by,
\begin{equation}
    \mathcal{D}=\frac{\mathcal{Y}_0^S}{\mathcal{Y}_0^C}\beta_R^C+\mathcal{Y}_1^S\left(\frac{\beta_R^C}{\mathcal{Y}_0^C}\right)'+\beta_R^S,
\end{equation}
where,
\begin{subequations}
    \begin{gather}
        \mathcal{Y}_0^C=\frac{4\bar{\iota}_0\kappa}{\eta}, \\
        \mathcal{Y}_0^S=\frac{4\kappa'-8\bar{\iota}_0\kappa\sigma}{\eta}, \\
        \mathcal{Y}_1^S=-\frac{4\kappa}{\eta}
    \end{gather}
    \begin{widetext}
    \begin{multline}
        \beta_R^C=\frac{1}{2 l'  \kappa}\left[4\eta^2B_{\alpha0}^2 \left(Y_{22}^C Z_{22}^S+Y_{22}^S \left(\tilde{Z}_{20}-Z_{22}^C\right)\right)+2 B_{\alpha0} \left(\eta ^2 Y_{22}^C \tau l'+\right.\right.\\
        \left.\left.+\eta ^2 {X_{22}^C}'
        -\kappa l'\tilde{Z}_{20} \left(\eta ^2-4 \kappa \left(X_{22}^C-\sigma X_{22}^S\right)\right)-4 l'\kappa^2 \sigma X_{22}^C Z_{22}^S-4 l'\kappa^2 \tilde{X}_{20} Z_{22}^C+\right.\right.\\
        \left.\left.+4l' \kappa^2 \sigma \tilde{X}_{20} Z_{22}^S+2 \eta ^2 \iota _0 X_{22}^S+4 l'\kappa^2 \sigma X_{22}^S Z_{22}^C+\eta ^2 \kappa l' Z_{22}^C-
        \eta ^2 \tilde{X}_{20}{}'\right)+\right.\\
        \left.+4\kappa^2 l' \left(\sigma \tilde{X}_{20} \tau l'- \sigma \tau l' X_{22}^C-\tau l' X_{22}^S+ \sigma \tilde{Y}_{22}^C{}'-2 \iota _0 (Y_{22}^C-\sigma Y_{22}^S)+ \tilde{Y}_{22}^S{}'\right)\right]
    \end{multline}
     \begin{multline}
        \beta_R^S=\frac{1}{2  \kappa l'}\left[2 B_{\alpha0} \left(\eta ^2 Y_{22}^S \tau l'-2 \eta ^2 \iota _0 X_{22}^C+4 \kappa^2 \sigma \tilde{Z}_{20} X_{22}^C l'-\right.\right.\\
        \left.\left.-4 \kappa^2 X_{22}^C Z_{22}^S-4 \kappa^2 \sigma \tilde{X}_{20} Z_{22}^C l'-4 \kappa^2 \tilde{X}_{20} Z_{22}^S l'+\eta ^2 {X_{22}^S}'+4 \kappa^2 X_{22}^S Z_{22}^C l'+\right.\right.\\
        \left.\left.+4 \kappa^2 \tilde{Z}_{20} X_{22}^S l'+\eta ^2 \kappa Z_{22}^S l'\right)+\kappa l' \left(4 \kappa \tilde{X}_{20} \tau l'+4 \kappa l' \tau X_{22}^C-4 \kappa l' \sigma \tau X_{22}^S-4 \kappa \tilde{Y}_{22}^C{}'-8 \iota _0 \kappa (\sigma \tilde{Y}_{22}^C+Y_{22}^S)+\right.\right.\\
        \left.\left.+4 \kappa \sigma \tilde{Y}_{22}^S{}'\right)-4 B_{\alpha0}^2 \eta ^2 \tilde{Y}_{22}^C \tilde{Z}_{20}\right],
    \end{multline}
   
\end{widetext}
\end{subequations}
The various forms of $X$, $Y$, and $Z$ are needed to complete the expression for $D$ (remembering that for the expressions we should make $B_{20}=0$ as we have dealt with these explicitly already). can be found explicitly in [\onlinecite{rodriguez2020i}] in the simplifying vacuum limit. More explicitly, the forms of $X$ are found in Eqs.~(14)-(15) and Appendix C; the expressions for $Y$ in Eqs.~(27)-(28); and the expressions for $Z$ in Eq.~(24) and following. For a systematic way to obtain these expressions, see [\onlinecite{rodriguez2021weak}]. This leaves a contribution to $D$ proportional to $B_{22}^C$. 

\section{Solution existence for $B_{20}$} \label{sec:appB20Schrod}

In this Appendix, we consider some existence and uniqueness properties of the $B_{20}$ equation, Eq.~(\ref{eqn:B20Eq}). We do so by assessing the second-order differential equation by applying the Fredholm Alternative theorem. 
\par
\hrulefill \par
\textbf{Fredholm Alternative Theorem}\label{def:FredAlt}
 \textit{   Let $\mathbb{L}$ be a linear operator with adjoint $\mathbb{L}^\dagger$. Then \textbf{exactly one} of the following is true:
    \begin{itemize}
        \item The inhomogeneous problem $\mathbb{L}y=f$ has a unique solution $y$.
        \item The homogeneous adjoint problem $\mathbb{L}^\dagger y=0$ has a non-trivial solution.
    \end{itemize}   
    In the event of the latter, the inhomogeneous equation has either no solution or infinitely many. If the solvability condition $\langle y_0,f\rangle=0$, then there are an infinite number of solutions.
}\par
\hrulefill
\par
For application of the Fredholm Alternative theorem, write Eq.~(\ref{eqn:B20Eq}) as,
\begin{equation}
    \mathbb{L}\frac{B_{20}}{B_0}+\mathcal{D}=[\mathcal{A}\partial_\phi^2+\mathcal{B}\partial_\phi+\mathcal{C}]\frac{B_{20}}{B_0}+\mathcal{D}=0, \label{eqn:B20eqL}
\end{equation} 
and complete the problem with periodic boundary conditions on $B_{20}$. 
\par
It is thus clear that we first need to study the homogeneous operator $\mathbb{L}$. To construct the adjoint of $\mathbb{L}$, it is convenient to rewrite Eq.~(\ref{eqn:B20eqL}) in a self-adjoint form. This can be achieved by writing the second-order ODE in the form of a Hill equation,
\begin{equation}
    \psi''+Q\psi=-\frac{D}{\mathcal{A}}e^{\int\frac{\mathcal{B}}{2\mathcal{A}}\mathrm{d}\phi}, \label{eqn:SchrodB20eq}
\end{equation}
where,
\begin{subequations}
    \begin{gather}
        Q=-\frac{1}{2}\left(\frac{\mathcal{B}}{\mathcal{A}}\right)'-\frac{1}{4}\left(\frac{\mathcal{B}}{\mathcal{A}}\right)^2+\frac{C}{\mathcal{A}}, \\
        \psi=B_{20}e^{-\int\left(\frac{\mathcal{B}}{2\mathcal{A}}\right)\mathrm{d}\phi}.
    \end{gather}
\end{subequations}
Using the coefficients $\mathcal{A}$ and $\mathcal{B}$, Eqs.~(\ref{eqn:AB20})-(\ref{eqn:BB20}),
\begin{equation*}
    \frac{\mathcal{B}}{\mathcal{A}}=2\bar{\iota}_0\sigma+\left[\ln\left(-\frac{\bar{\beta}_1^C}{\mathcal{Y}_1^S}\right)\right]', 
\end{equation*}
where $\mathcal{Y}_1^S=-4\kappa/\eta$ and $\bar{\beta}_1^C=B_{\alpha0}\eta^2 l^2\left[1+4\kappa^4(1+\sigma^2)/\eta^4B_{\alpha0}^2\right]/8\kappa^3\bar{\iota}_0$. The exponential factor that maps $B_{20}\rightarrow\psi$ is then by construction periodic. Therefore, solving the $B_{20}$ equation is equivalent to solving Eq.~(\ref{eqn:SchrodB20eq}) with periodic boundary conditions on $\psi$. This mapping preserves the even parity of $B_{20}$ in stellarator symmetry, and thus we may focus on Eq.~(\ref{eqn:SchrodB20eq}).  
\par
As the homogeneous operator of Eq.~(\ref{eqn:SchrodB20eq}) is self-adjoint, the adjoint problem for the Fredholm Alternative is $\psi''+Q\psi=0$. The problem is nothing but a periodic, time-independent Schr\"{o}dinger equation. With this interpretation, the `quantum potential' is $V=-Q/2$ (formally taking $\hbar,m=1$). Due to periodicity, the potential can be thought to represent a `crystal' (or periodic lattice) of period $\Delta\phi=2\pi/N$. The structure of this potential is determined by the axis shape and choice of $\eta$. Whenever this adjoint periodic Schr\"{o}dinger equation supports a zero energy state, then by the Fredholm Alternative, there will be no unique solution to the original $B_{20}$ equation. This may mean that no solution or an infinite number of them exists, a distinction in which the second-order choices will intervene through $\mathcal{D}$. 
\par
\begin{figure*}
    \centering
    \includegraphics[width=\textwidth]{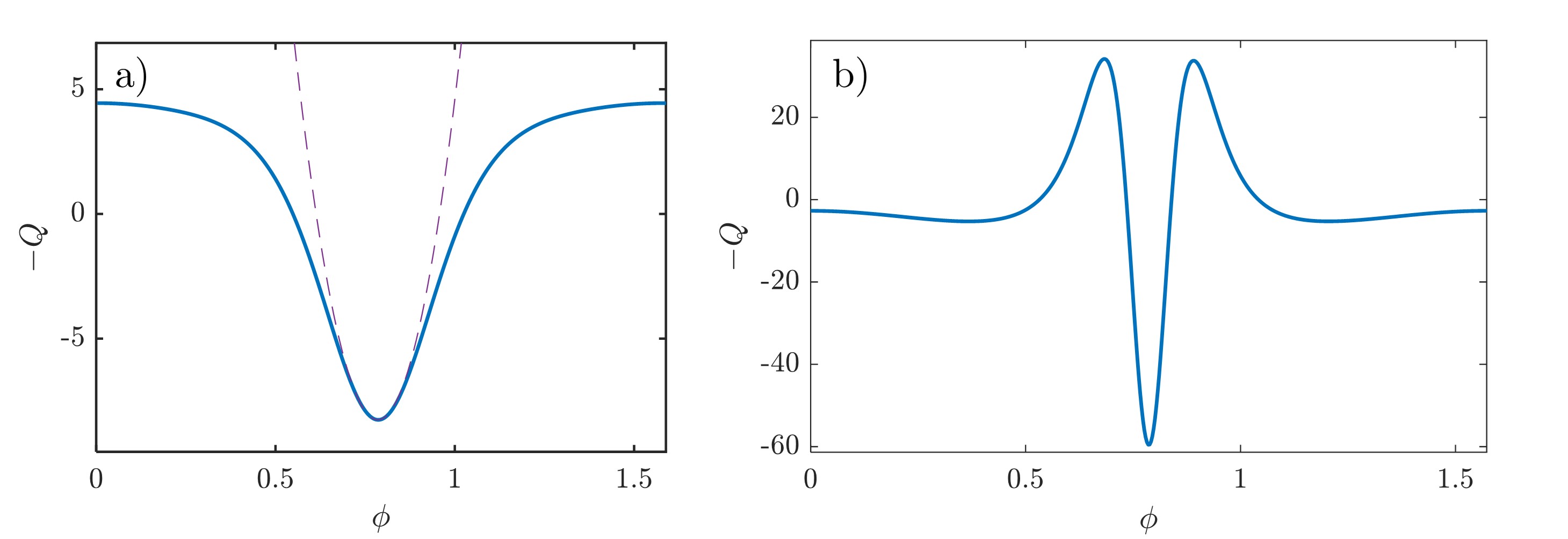}
    \caption{\textbf{Example of effective quantum potential for the adjoint problem.} This shows examples of the potential $-Q/2=V$ for (a) a QA with $R=1+0.0144\cos4\phi$, $Z=0.0144\sin4\phi$, $N=4$ and $\eta=2.1$ and (b) a QH with $R=1+0.1\cos4\phi$, $Z=0.1\sin4\phi$, $N=4$ and $\eta=2$. The broken line in (a) shows a quadratic potential centred around the minimum. These are two examples of what the effective potential wells of the adjoint problem look like.} 
    \label{fig:exampleB20pot}
\end{figure*}

For a generally shaped potential (see Fig.~\ref{fig:exampleB20pot}), there is no closed form for the energy eigenstates of the adjoint. However, the discrete nature of energy eigenvalues and the very special requirement for a vanishing energy state suggests that, in general, such an eigenstate will not exist. Thus, a solution to the direct $B_{20}$ problem will. Only for very particular choices will the zero energy eigenstate exist, leading to the divergence observed in Fig.~\ref{fig:singB20example}. Such special value for a fixed axis shape corresponds to $\eta_\mathrm{crit}$. We shall illustrate these abstract statements with an example in which the existence or not of a solution is considered.
 \par
For example, the near-axis construction corresponding to Fig.~\ref{fig:exampleB20pot}, in which the potential has a simple finite depth and width well in each lattice cell. In general, more complex features will be present, as also shown in the figure. Naturally, the first guess to the form of the eigenstates are states corresponding to single finite depth well, a standard textbook problem [\onlinecite[Chap.~2.6]{griffiths2004QM}]. Of course, the solution will be significantly more involved than this. For once, in a lattice, wavefunctions are generally not isolated to each cell, but there is a hopping energy and an overlap between wavefunctions in nearby lattice sites that lead to the splitting of the energy states. Secondly, the shapes of the potential wells change the eigenstructure of the problem. Of course, not having an exact solution, we would like to estimate the energy spectrum of the problem without depending excessively on the model used. This is achieved through a variational approach in quantum mechanics. From the orthogonality of the energy eigenstates and focusing on the ground-state energy, this can be estimated by considering
\begin{equation}
    E_0\leq \frac{\int\left[(\psi')^2-Q\psi^2\right]\mathrm{d}\phi}{\int\psi^2\mathrm{d}\phi}, \label{eqn:GSenerEst}
\end{equation}
where $\psi$ can be any (here real) function. To get a tighter bound on the energy, an informed guess of $\psi$ is needed. Leaving some parameters of $\psi$ as unknowns (say concatenated single-well wavefunctions with free depth and width parameters), a minimum to the expression may be sought, with the resulting value being an approximation to the energy state. 
\begin{figure}
    \centering
    \includegraphics[width=0.5\textwidth]{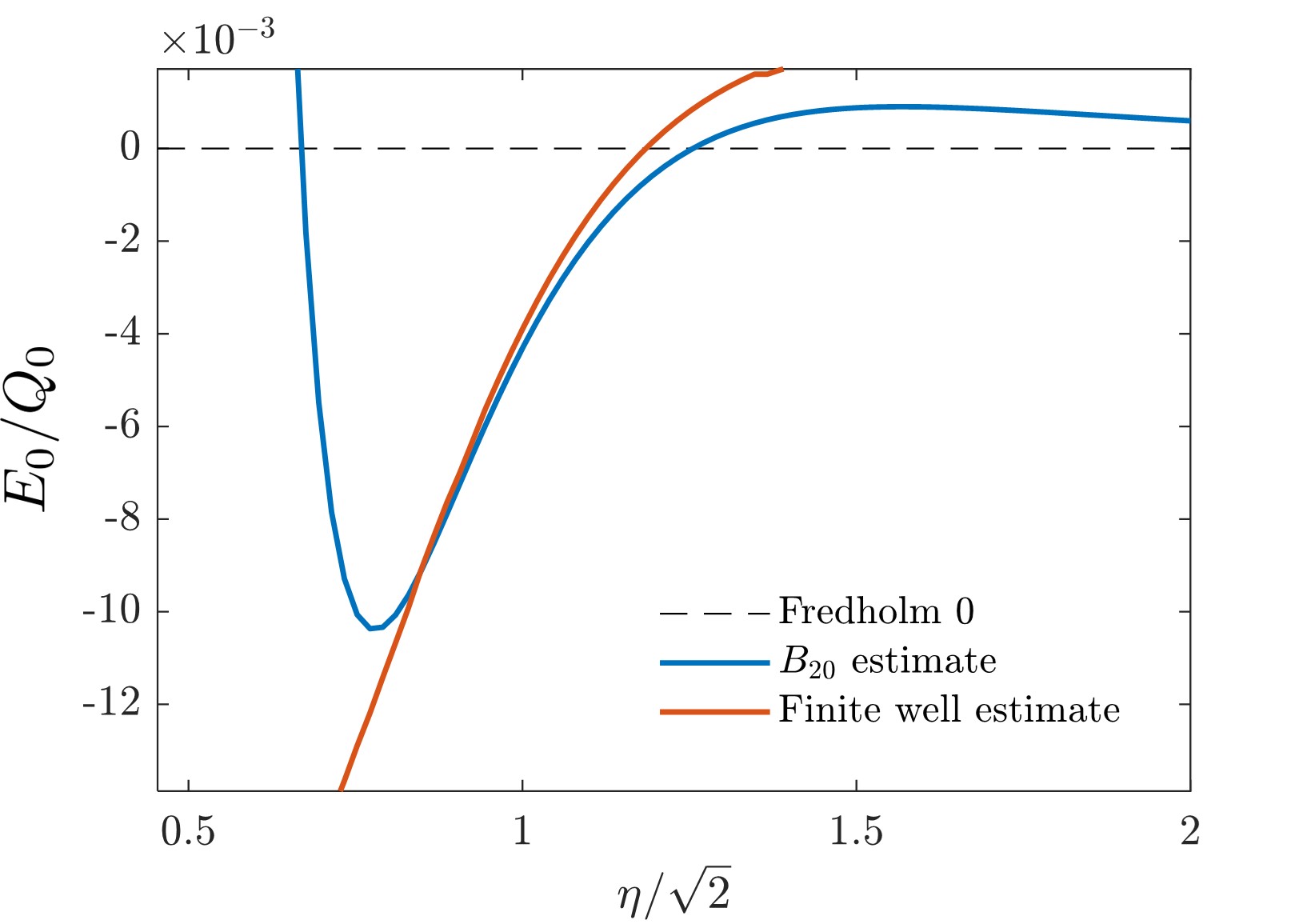}
    \caption{\textbf{Ground-state energy of adjoint problem as a function of $\eta$.} The plot shows the evaluation of the adjoint ground-state energies ($E_0$ normalised to $Q$ at the bottom of the potential troughs, $Q_0$) as a function of the $\eta$ parameter for the QA example in Fig.~\ref{fig:exampleB20pot}. The blue curve shows the estimate of the ground-state using the wavefunction guess from $B_{20}$, and the orange shows the finite well variational approach. The estimated ground-state can be considered the smallest of these two curves. The crossing with the 0 for the $B_{20}$ estimate is precisely the point at which the resonance occurs in Fig.~\ref{fig:exampleB20pot}. }
    \label{fig:B20potRes}
\end{figure}
\par
For most values of $\eta$ (see the domain in which $E_0<0$ in Fig.~\ref{fig:B20potRes}), the ground-state lies below the zero energy level, and the $B_{20}$ solution is unique. This assumes that none of the higher energy states resonate, which is true given the proximity of the energy $E_0$ to 0 (compared to the well depth) and the parity requirement on the solution, which makes the next energy level solution lie significantly higher. Although we do not prove it here, we shall assume that only the ground-state is relevant here. The precise location of the 0-energy crossing will vary with $\psi_\mathrm{guess}$, and thus only a lower bound may be given for $\eta_\mathrm{crit}$. A more precise crossing value can be obtained by constructing $\psi$ from the solution of $B_{20}$. This is generally not a great guess for the adjoint problem, except in the region close to where the existence of the solution starts to break down: the critical $\eta$ value. If this critical value indicates the second alternative of the Fredholm Alternative, then in the neighbourhood of this critical resonance, the solution $B_{20}$ will be dominated by the ground-state solution of the adjoint problem, and thus the zero crossing will be exact (see Fig.~\ref{fig:B20potRes}). At this point, the choice of $B_{22}^C$ is important, as it determines whether the inhomogeneous term in Eq.~(\ref{eqn:SchrodB20eq}) is orthogonal to the ground-state solution or not, controlling whether there exist none or an \textit{infinite} number of solutions, where the infinite family would correspond to the addition of an arbitrary multiple of the ground-state $\psi_0$.  
\par 
Solutions do exist in general and are unique, except for particular critical values of $\eta$ for which the ground-state of the adjoint problem has vanishing energy. 
\begin{figure}
    \centering
    \includegraphics[width=0.2\textwidth]{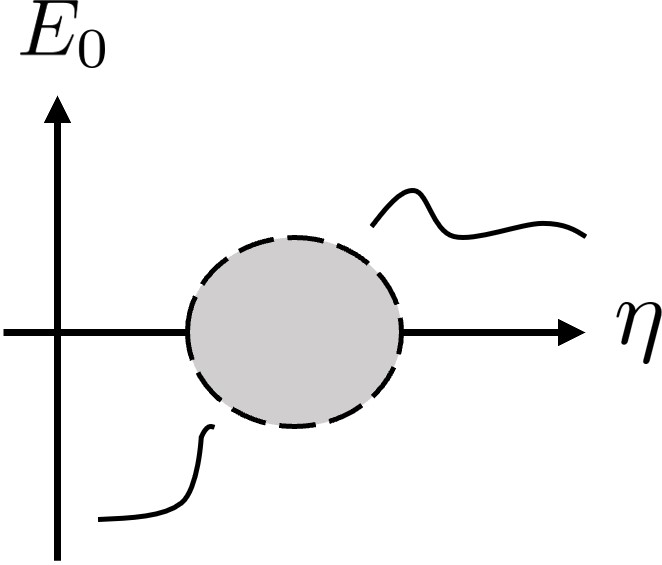}
    \caption{\textbf{Diagram describing the idea behind the existence proof.} Illustration of the rationale behind the proof of existence of a singular value $\eta_\mathrm{crit}$ to the $B_{20}$ equation.}
    \label{fig:singB20diagrProof}
\end{figure}
The existence of such a value has not been proven yet. We do so now by showing that the energy state of the adjoint homogeneous equation necessarily crosses the zero. Assuming continuity with $\eta$, it will suffice to show that the energy level is positive (negative) for small (large) $\eta$ (see Fig.~\ref{fig:singB20diagrProof}). 
\par
Let us start with the small $\eta$ limit, for which $\mathcal{A}\sim O(1/\eta^5)$, $\mathcal{B}\sim O(1/\eta)$ and $\mathcal{C}\sim O(1/\eta)$, from Eqs.~(\ref{eqn:AB20})-(\ref{eqn:CB20}). To leading order $O(\eta^4)$ then $Q\sim-(\mathcal{B}/\mathcal{A})'/2+\mathcal{C}/\mathcal{A}$. In this limit, it is sufficient to show that the estimate of the ground-state energy is negative. Choose the wavefunction to be constant, $\psi=\psi_0$. Then, 
\begin{equation*}
    E_0\leq \frac{\int_0^{2\pi}\left[(\psi')^2-Q\psi^2\right]\mathrm{d}\phi}{\int_0^{2\pi}\psi^2\mathrm{d}\phi}=-\frac{1}{2\pi}\int_0^{2\pi} Q\mathrm{d}\phi.
\end{equation*}
Given the asymptotic form of $Q$, 
\begin{equation*}
    \int_0^{2\pi} Q\mathrm{d}\phi=\int_0^{2\pi}\frac{\mathcal{C}}{\mathcal{A}}\mathrm{d}\phi\sim\frac{3\bar{\iota}_0^2}{B_{\alpha0}(l')^2}2\pi\geq 0.
\end{equation*}
Thus, for a small (but finite $\eta$), the energy state of the adjoint homogeneous equation has ground-state energy $E_0<0$. 
\par
In the large $\eta$ limit, $\mathcal{A}\sim O(\eta^3)$, $\mathcal{B}\sim O(\eta^3)$ and $\mathcal{C}\sim O(1/\eta)$. Thus, the potential $Q\sim-(\mathcal{B}/\mathcal{A})'/2-(\mathcal{B}/\mathcal{A})^2/4$ becomes independent of $\mathcal{C}$ allowing one to rewrite the adjoint problem defining $y=\psi \exp[-\int (\mathcal{B}/2\mathcal{A})\mathrm{d}\phi]$, 
\begin{equation}
    y''+\frac{\mathcal{B}}{\mathcal{A}}y'\sim0,
\end{equation}
which can be solved exactly by $y\sim\mathbb{C}\int\exp[-\int(\mathcal{B}/\mathcal{A})\mathrm{d}\phi']\mathrm{d}\phi$. The exponential must be positive; thus, its integral is non-periodic, making $y$ non-periodic as well. Thus, the only solution is the trivial $y=0$. Due to the adjoint problem only having a trivial solution, in the large $\eta$ limit, the solution to the $B_{20}$ equation is once again unique. However, this does not provide us with the ground-state energy of the adjoint equation. To learn what that is, write $y''+(\mathcal{B}/\mathcal{A})y'\sim-E_0y$, where we are explicitly including the energy eigenvalue associated with the ground-state. We multiply the equation by $y$ and integrate over $\phi$, so that 
\begin{equation*}
    y\left[Ey'\right]'\sim-E_0Ey^2\rightarrow E_0\sim\frac{\int(y')^2E\mathrm{d}\phi}{\int y^2E\mathrm{d}\phi}>0.
\end{equation*} 
Here $E=\exp[2\bar{\iota}_0\int\sigma\mathrm{d}\phi]$. Regardless of the choice of $y$, the ground-state is necessarily positive. Thus, as the ground-state energy of the adjoint problem in these two limits has opposite signs, by the mean-value theorem, it must cross the zero at some value of $\eta=\eta_\mathrm{crit}$. This proof does not guarantee the uniqueness of this singularity (see Fig.~\ref{fig:singB20example}), but it guarantees its presence. Given the numerical evidence, we shall, for the remainder of this Appendix, assume it to be unique.
\par
All of this has introduced a new special $\eta$ value, $\eta_\mathrm{crit}$, which appears not to bear any relation to the choice of $\eta^*$ introduced in the main text. We would like to relate the two. For a unique $\eta_\mathrm{crit}$, the sign of $E_0$ at a given $\eta$ determines its position relative to $\eta_\mathrm{crit}$: $E_0<0$ for $\eta<\eta_\mathrm{crit}$ and $E_0>0$ for $\eta>\eta_\mathrm{crit}$. Thus, we would like to assess the sign of the ground-state energy associated with $\eta^*$ to situate it to the right or left of $\eta_\mathrm{crit}$. 
\par
Let us again exploit the Ritz variational form of energy, Eq.~(\ref{eqn:GSenerEst}), in this case using a trial wavefunction $\psi\sim\exp\left[\int(\mathcal{B}/2\mathcal{A})\mathrm{d}\phi\right]$. A wavefunction of this form is suggested by the defining condition of $\eta^*$, Eq.~(\ref{eqn:extrEtaIotaRel}). For the ground-state energy threshold, the following are required, 
\begin{gather*}
    \int\left(\frac{\mathcal{B}}{\mathcal{A}}\right)^2\psi^2\mathrm{d}\phi=4\int(\psi')^2\mathrm{d}\phi, \\
    \int\left(\frac{\mathcal{B}}{\mathcal{A}}\right)'\psi^2\mathrm{d}\phi=-4\int(\psi')^2\mathrm{d}\phi, \\
    \int \frac{\mathcal{C}}{\mathcal{A}}\psi^2\mathrm{d}\phi=\frac{|\bar{\iota}_0|}{4B_{\alpha0}\eta^*}\int e^{2\bar{\iota}_0\int\sigma\mathrm{d}\phi'}\mathrm{d}\phi.
\end{gather*}  
For the latter, the forms of $\mathcal{A}$ and $\mathcal{B}$ in Eqs.~(\ref{eqn:AB20})-(\ref{eqn:BB20}) were used, as well as integration by parts using Eq.~(\ref{eqn:extrEtaIotaRel}). Putting them together,
\begin{equation}
    \int\left[(\psi')^2-Q\psi^2\right]\mathrm{d}\phi=-\int\frac{\mathcal{C}}{\mathcal{A}}\psi^2\mathrm{d}\phi<0,
\end{equation}
which thus yields a ground-state energy $E_0(\eta^*)<0$. Thus, it follows that $\eta^*<\eta_\mathrm{crit}$. The choice of $\eta^*$ thus naturally avoids the divergence of the $B_{20}$ equation, as will a search in $0<\eta<\eta^*$. This observation may be taken as additional evidence in favour of the choice of $\eta^*$ first order.   

\section{Bounding $B_{22}^C$ search} \label{sec:appB2cBound}
We argued in the main text that to avoid extreme second-order shaping, we should put some bounds on the allowable $B_{22}^C$. Following [\onlinecite{rodriguez2023mhd}], it is straightforward to see that both the Shafranov shift as well as the standard measure of triangularity increase linearly with $B_{22}^C$ in the large $|B_{22}^C|$ limit. Furthermore, with this dependence, the construction becomes unphysical if $|B_{22}^C|$ is too large. 
\par
To make things more quantitative, consider an up-down symmetric cross-section in the stellarator as representative of the behaviour of shaping in the configuration, and focus on the behaviour of its Shafranov shift. This describes the relative displacement of the centres of cross-sections when going from one flux surface to the next. If the relative displacement of the centres of the cross-sections at different $\psi$ is too large, then eventually flux surfaces will intersect each other (this phenomenon of flux surface intersection gives rise to the measure $r_c$, the largest value of $\epsilon$ in which the near-axis description is sensible, presented in [\onlinecite{landreman2021a}]). This is unacceptable and thus can be leveraged to estimate a bound on $B_{22}^C$. We reproduce from [\onlinecite{rodriguez2022mhd}] a simple estimate of when this situation is reached.
\par
Start for simplicity with a second-order construction in which, besides the elliptic shape, the second-order shaping has $X_2=X_{20}+X_{22}^C\cos2\chi$. Focus then on the centre-line of the cross-section, about which it is up-down symmetric. Looking at $\chi=0$, $X=\epsilon X_{11}^C+\epsilon^2(X_{20}+X_{22}^C)$. Now the intersection occurs whenever $\partial_\epsilon X=0$, which can be solved for the critical Shafranov shift,
\begin{equation}
    X_{20}+X_{22}^C=\frac{X_{11}^C}{2\epsilon}.
\end{equation}
Whenever the Shafranov shift exceeds the value on the RHS, then the cross-sections at the stellarator symmetric point will intersect. 
\par
Let us see how a large $|B_{22}^C|$ affects the Shafranov shift. To do that, we need to learn about the behaviour of $X_{20}$ and $X_{22}^C$ at $\phi=0$. These expressions may be found by careful consideration of the symmetry properties of the various functions involved. As a result,
\begin{equation}
    X_{20}+X_{22}^C\sim-\frac{1}{2\kappa}\frac{B_{22}^C}{B_0}\left(1+\frac{B_{20}^\mathrm{univ}}{B_0}\right).
\end{equation}
Thus, a critical $B_{22}^C$,
\begin{equation}
    \frac{|B_{22}^C|}{B_0}\sim\frac{\kappa X_{11}^C}{\epsilon}\left(1+\frac{B_{20}^\mathrm{univ}}{B_0}\right)^{-1}.
\end{equation}
Crudely, taking $\eta=\kappa X_{11}^C\sim 1$ and ignoring the expression in the bracket, we get $|B_{22}^C|/B_0\sim1/\epsilon$. That is, we may roughly consider it proportional to the `aspect ratio,' $1/\epsilon$, of the configuration. With a reasonable aspect ratio of $1/\epsilon\sim10$, we obtain the $B_{22}^C$ limit in the text (which could be relaxed by allowing for other larger values like, e.g., 20).  
\par
It should be clear from this approach that the estimate is but a crude one, yet nevertheless useful. For instance, we are ignoring $B_{20}^\mathrm{univ}$, which could relax this bound significantly depending on the situation. In practice, the construction of this crude bound allows us to perform bounded optimisation. In most of the relevant space (where the most reasonably shaped, quasisymmetric configurations lie), it has not had much of an effect on the result (the $B_{22}^C$ minimum is well within the interval, as an example, see Fig.~\ref{fig:B22cOptpreciseQS}).

\bibliography{topSpaceQS}

\begin{thebibliography}{63}%
\makeatletter
\providecommand \@ifxundefined [1]{%
 \@ifx{#1\undefined}
}%
\providecommand \@ifnum [1]{%
 \ifnum #1\expandafter \@firstoftwo
 \else \expandafter \@secondoftwo
 \fi
}%
\providecommand \@ifx [1]{%
 \ifx #1\expandafter \@firstoftwo
 \else \expandafter \@secondoftwo
 \fi
}%
\providecommand \natexlab [1]{#1}%
\providecommand \enquote  [1]{``#1''}%
\providecommand \bibnamefont  [1]{#1}%
\providecommand \bibfnamefont [1]{#1}%
\providecommand \citenamefont [1]{#1}%
\providecommand \href@noop [0]{\@secondoftwo}%
\providecommand \href [0]{\begingroup \@sanitize@url \@href}%
\providecommand \@href[1]{\@@startlink{#1}\@@href}%
\providecommand \@@href[1]{\endgroup#1\@@endlink}%
\providecommand \@sanitize@url [0]{\catcode `\\12\catcode `\$12\catcode
  `\&12\catcode `\#12\catcode `\^12\catcode `\_12\catcode `\%12\relax}%
\providecommand \@@startlink[1]{}%
\providecommand \@@endlink[0]{}%
\providecommand \url  [0]{\begingroup\@sanitize@url \@url }%
\providecommand \@url [1]{\endgroup\@href {#1}{\urlprefix }}%
\providecommand \urlprefix  [0]{URL }%
\providecommand \Eprint [0]{\href }%
\providecommand \doibase [0]{https://doi.org/}%
\providecommand \selectlanguage [0]{\@gobble}%
\providecommand \bibinfo  [0]{\@secondoftwo}%
\providecommand \bibfield  [0]{\@secondoftwo}%
\providecommand \translation [1]{[#1]}%
\providecommand \BibitemOpen [0]{}%
\providecommand \bibitemStop [0]{}%
\providecommand \bibitemNoStop [0]{.\EOS\space}%
\providecommand \EOS [0]{\spacefactor3000\relax}%
\providecommand \BibitemShut  [1]{\csname bibitem#1\endcsname}%
\let\auto@bib@innerbib\@empty
\bibitem [{\citenamefont {Bernardin}, \citenamefont {Moses},\ and\
  \citenamefont {Tataronis}(1986)}]{bernardin1986}%
  \BibitemOpen
  \bibfield  {author} {\bibinfo {author} {\bibfnamefont {M.~P.}\ \bibnamefont
  {Bernardin}}, \bibinfo {author} {\bibfnamefont {R.~W.}\ \bibnamefont
  {Moses}},\ and\ \bibinfo {author} {\bibfnamefont {J.~A.}\ \bibnamefont
  {Tataronis}},\ }\bibfield  {title} {\enquote {\bibinfo {title} {Isodynamical
  (omnigenous) equilibrium in symmetrically confined plasma configurations},}\
  }\href {https://doi.org/10.1063/1.865501} {\bibfield  {journal} {\bibinfo
  {journal} {The Physics of Fluids}\ }\textbf {\bibinfo {volume} {29}},\
  \bibinfo {pages} {2605--2611} (\bibinfo {year} {1986})}\BibitemShut {NoStop}%
\bibitem [{\citenamefont {Cary}\ and\ \citenamefont
  {Shasharina}(1997)}]{cary1997}%
  \BibitemOpen
  \bibfield  {author} {\bibinfo {author} {\bibfnamefont {J.~R.}\ \bibnamefont
  {Cary}}\ and\ \bibinfo {author} {\bibfnamefont {S.~G.}\ \bibnamefont
  {Shasharina}},\ }\bibfield  {title} {\enquote {\bibinfo {title} {Omnigenity
  and quasihelicity in helical plasma confinement systems},}\ }\href
  {https://doi.org/10.1063/1.872473} {\bibfield  {journal} {\bibinfo  {journal}
  {Physics of Plasmas}\ }\textbf {\bibinfo {volume} {4}},\ \bibinfo {pages}
  {3323--3333} (\bibinfo {year} {1997})}\BibitemShut {NoStop}%
\bibitem [{\citenamefont {Hall}\ and\ \citenamefont
  {McNamara}(1975)}]{hall1975}%
  \BibitemOpen
  \bibfield  {author} {\bibinfo {author} {\bibfnamefont {L.~S.}\ \bibnamefont
  {Hall}}\ and\ \bibinfo {author} {\bibfnamefont {B.}~\bibnamefont
  {McNamara}},\ }\bibfield  {title} {\enquote {\bibinfo {title}
  {Three‐dimensional equilibrium of the anisotropic, finite‐pressure
  guiding‐center plasma: Theory of the magnetic plasma},}\ }\href@noop {}
  {\bibfield  {journal} {\bibinfo  {journal} {The Physics of Fluids}\ }\textbf
  {\bibinfo {volume} {18}},\ \bibinfo {pages} {552--565} (\bibinfo {year}
  {1975})}\BibitemShut {NoStop}%
\bibitem [{\citenamefont {Landreman}\ and\ \citenamefont
  {Catto}(2012)}]{landreman2012}%
  \BibitemOpen
  \bibfield  {author} {\bibinfo {author} {\bibfnamefont {M.}~\bibnamefont
  {Landreman}}\ and\ \bibinfo {author} {\bibfnamefont {P.~J.}\ \bibnamefont
  {Catto}},\ }\bibfield  {title} {\enquote {\bibinfo {title} {Omnigenity as
  generalized quasisymmetry},}\ }\href {https://doi.org/10.1063/1.3693187}
  {\bibfield  {journal} {\bibinfo  {journal} {Physics of Plasmas}\ }\textbf
  {\bibinfo {volume} {19}},\ \bibinfo {pages} {056103} (\bibinfo {year}
  {2012})}\BibitemShut {NoStop}%
\bibitem [{\citenamefont {Helander}(2014)}]{Helander2014}%
  \BibitemOpen
  \bibfield  {author} {\bibinfo {author} {\bibfnamefont {P.}~\bibnamefont
  {Helander}},\ }\bibfield  {title} {\enquote {\bibinfo {title} {Theory of
  plasma confinement in non-axisymmetric magnetic fields},}\ }\href
  {https://doi.org/10.1088/0034-4885/77/8/087001} {\bibfield  {journal}
  {\bibinfo  {journal} {Reports on Progress in Physics}\ }\textbf {\bibinfo
  {volume} {77}},\ \bibinfo {pages} {087001} (\bibinfo {year}
  {2014})}\BibitemShut {NoStop}%
\bibitem [{\citenamefont {Boozer}(1983)}]{boozer1983}%
  \BibitemOpen
  \bibfield  {author} {\bibinfo {author} {\bibfnamefont {A.~H.}\ \bibnamefont
  {Boozer}},\ }\bibfield  {title} {\enquote {\bibinfo {title} {Transport and
  isomorphic equilibria},}\ }\href@noop {} {\bibfield  {journal} {\bibinfo
  {journal} {The Physics of Fluids}\ }\textbf {\bibinfo {volume} {26}},\
  \bibinfo {pages} {496--499} (\bibinfo {year} {1983})}\BibitemShut {NoStop}%
\bibitem [{\citenamefont {N{\"u}hrenberg}\ and\ \citenamefont
  {Zille}(1988)}]{nuhren1988}%
  \BibitemOpen
  \bibfield  {author} {\bibinfo {author} {\bibfnamefont {J.}~\bibnamefont
  {N{\"u}hrenberg}}\ and\ \bibinfo {author} {\bibfnamefont {R.}~\bibnamefont
  {Zille}},\ }\bibfield  {title} {\enquote {\bibinfo {title} {Quasi-helically
  symmetric toroidal stellarators},}\ }\href@noop {} {\bibfield  {journal}
  {\bibinfo  {journal} {Physics Letters A}\ }\textbf {\bibinfo {volume}
  {129}},\ \bibinfo {pages} {113 -- 117} (\bibinfo {year} {1988})}\BibitemShut
  {NoStop}%
\bibitem [{\citenamefont {Rodríguez}, \citenamefont {Helander},\ and\
  \citenamefont {Bhattacharjee}(2020)}]{rodriguez2020}%
  \BibitemOpen
  \bibfield  {author} {\bibinfo {author} {\bibfnamefont {E.}~\bibnamefont
  {Rodríguez}}, \bibinfo {author} {\bibfnamefont {P.}~\bibnamefont
  {Helander}},\ and\ \bibinfo {author} {\bibfnamefont {A.}~\bibnamefont
  {Bhattacharjee}},\ }\bibfield  {title} {\enquote {\bibinfo {title} {Necessary
  and sufficient conditions for quasisymmetry},}\ }\href
  {https://doi.org/10.1063/5.0008551} {\bibfield  {journal} {\bibinfo
  {journal} {Physics of Plasmas}\ }\textbf {\bibinfo {volume} {27}},\ \bibinfo
  {pages} {062501} (\bibinfo {year} {2020})}\BibitemShut {NoStop}%
\bibitem [{\citenamefont {Anderson}\ \emph {et~al.}(1995)\citenamefont
  {Anderson}, \citenamefont {Almagri}, \citenamefont {Anderson}, \citenamefont
  {Matthews}, \citenamefont {Talmadge},\ and\ \citenamefont
  {Shohet}}]{Anderson1995}%
  \BibitemOpen
  \bibfield  {author} {\bibinfo {author} {\bibfnamefont {F.~S.~B.}\
  \bibnamefont {Anderson}}, \bibinfo {author} {\bibfnamefont {A.~F.}\
  \bibnamefont {Almagri}}, \bibinfo {author} {\bibfnamefont {D.~T.}\
  \bibnamefont {Anderson}}, \bibinfo {author} {\bibfnamefont {P.~G.}\
  \bibnamefont {Matthews}}, \bibinfo {author} {\bibfnamefont {J.~N.}\
  \bibnamefont {Talmadge}},\ and\ \bibinfo {author} {\bibfnamefont {J.~L.}\
  \bibnamefont {Shohet}},\ }\bibfield  {title} {\enquote {\bibinfo {title} {The
  {Helically Symmetric Experiment (HSX)} goals, design and status},}\
  }\href@noop {} {\bibfield  {journal} {\bibinfo  {journal} {Fusion
  Technology}\ }\textbf {\bibinfo {volume} {27}},\ \bibinfo {pages} {273--277}
  (\bibinfo {year} {1995})}\BibitemShut {NoStop}%
\bibitem [{\citenamefont {Zarnstorff}\ \emph {et~al.}(2001)\citenamefont
  {Zarnstorff}, \citenamefont {Berry}, \citenamefont {Brooks}, \citenamefont
  {Fredrickson}, \citenamefont {Fu}, \citenamefont {Hirshman}, \citenamefont
  {Hudson}, \citenamefont {Ku}, \citenamefont {Lazarus}, \citenamefont
  {Mikkelsen} \emph {et~al.}}]{Zarnstorff2001}%
  \BibitemOpen
  \bibfield  {author} {\bibinfo {author} {\bibfnamefont {M.}~\bibnamefont
  {Zarnstorff}}, \bibinfo {author} {\bibfnamefont {L.}~\bibnamefont {Berry}},
  \bibinfo {author} {\bibfnamefont {A.}~\bibnamefont {Brooks}}, \bibinfo
  {author} {\bibfnamefont {E.}~\bibnamefont {Fredrickson}}, \bibinfo {author}
  {\bibfnamefont {G.}~\bibnamefont {Fu}}, \bibinfo {author} {\bibfnamefont
  {S.}~\bibnamefont {Hirshman}}, \bibinfo {author} {\bibfnamefont
  {S.}~\bibnamefont {Hudson}}, \bibinfo {author} {\bibfnamefont
  {L.}~\bibnamefont {Ku}}, \bibinfo {author} {\bibfnamefont {E.}~\bibnamefont
  {Lazarus}}, \bibinfo {author} {\bibfnamefont {D.}~\bibnamefont {Mikkelsen}},
  \emph {et~al.},\ }\bibfield  {title} {\enquote {\bibinfo {title} {Physics of
  the compact advanced stellarator ncsx},}\ }\href@noop {} {\bibfield
  {journal} {\bibinfo  {journal} {Plasma Physics and Controlled Fusion}\
  }\textbf {\bibinfo {volume} {43}},\ \bibinfo {pages} {A237} (\bibinfo {year}
  {2001})}\BibitemShut {NoStop}%
\bibitem [{\citenamefont {Najmabadi}\ \emph {et~al.}(2008)\citenamefont
  {Najmabadi}, \citenamefont {Raffray}, \citenamefont {Abdel-Khalik},
  \citenamefont {Bromberg}, \citenamefont {Crosatti}, \citenamefont
  {El-Guebaly}, \citenamefont {Garabedian}, \citenamefont {Grossman},
  \citenamefont {Henderson}, \citenamefont {Ibrahim} \emph
  {et~al.}}]{Najmabadi2008}%
  \BibitemOpen
  \bibfield  {author} {\bibinfo {author} {\bibfnamefont {F.}~\bibnamefont
  {Najmabadi}}, \bibinfo {author} {\bibfnamefont {A.}~\bibnamefont {Raffray}},
  \bibinfo {author} {\bibfnamefont {S.}~\bibnamefont {Abdel-Khalik}}, \bibinfo
  {author} {\bibfnamefont {L.}~\bibnamefont {Bromberg}}, \bibinfo {author}
  {\bibfnamefont {L.}~\bibnamefont {Crosatti}}, \bibinfo {author}
  {\bibfnamefont {L.}~\bibnamefont {El-Guebaly}}, \bibinfo {author}
  {\bibfnamefont {P.}~\bibnamefont {Garabedian}}, \bibinfo {author}
  {\bibfnamefont {A.}~\bibnamefont {Grossman}}, \bibinfo {author}
  {\bibfnamefont {D.}~\bibnamefont {Henderson}}, \bibinfo {author}
  {\bibfnamefont {A.}~\bibnamefont {Ibrahim}}, \emph {et~al.},\ }\bibfield
  {title} {\enquote {\bibinfo {title} {The {ARIES-CS} compact stellarator
  fusion power plant},}\ }\href@noop {} {\bibfield  {journal} {\bibinfo
  {journal} {Fusion Science and Technology}\ }\textbf {\bibinfo {volume}
  {54}},\ \bibinfo {pages} {655--672} (\bibinfo {year} {2008})}\BibitemShut
  {NoStop}%
\bibitem [{\citenamefont {Ku}\ and\ \citenamefont {Boozer}(2010)}]{Ku2010}%
  \BibitemOpen
  \bibfield  {author} {\bibinfo {author} {\bibfnamefont {L.}~\bibnamefont
  {Ku}}\ and\ \bibinfo {author} {\bibfnamefont {A.}~\bibnamefont {Boozer}},\
  }\bibfield  {title} {\enquote {\bibinfo {title} {New classes of
  quasi-helically symmetric stellarators},}\ }\href@noop {} {\bibfield
  {journal} {\bibinfo  {journal} {Nuclear Fusion}\ }\textbf {\bibinfo {volume}
  {51}},\ \bibinfo {pages} {013004} (\bibinfo {year} {2010})}\BibitemShut
  {NoStop}%
\bibitem [{\citenamefont {Bader}\ \emph {et~al.}(2019)\citenamefont {Bader},
  \citenamefont {Drevlak}, \citenamefont {Anderson}, \citenamefont {Faber},
  \citenamefont {Hegna}, \citenamefont {Likin}, \citenamefont {Schmitt},\ and\
  \citenamefont {Talmadge}}]{bader2019}%
  \BibitemOpen
  \bibfield  {author} {\bibinfo {author} {\bibfnamefont {A.}~\bibnamefont
  {Bader}}, \bibinfo {author} {\bibfnamefont {M.}~\bibnamefont {Drevlak}},
  \bibinfo {author} {\bibfnamefont {D.~T.}\ \bibnamefont {Anderson}}, \bibinfo
  {author} {\bibfnamefont {B.~J.}\ \bibnamefont {Faber}}, \bibinfo {author}
  {\bibfnamefont {C.~C.}\ \bibnamefont {Hegna}}, \bibinfo {author}
  {\bibfnamefont {K.~M.}\ \bibnamefont {Likin}}, \bibinfo {author}
  {\bibfnamefont {J.~C.}\ \bibnamefont {Schmitt}},\ and\ \bibinfo {author}
  {\bibfnamefont {J.~N.}\ \bibnamefont {Talmadge}},\ }\bibfield  {title}
  {\enquote {\bibinfo {title} {Stellarator equilibria with reactor relevant
  energetic particle losses},}\ }\href
  {https://doi.org/10.1017/S0022377819000680} {\bibfield  {journal} {\bibinfo
  {journal} {Journal of Plasma Physics}\ }\textbf {\bibinfo {volume} {85}},\
  \bibinfo {pages} {905850508} (\bibinfo {year} {2019})}\BibitemShut {NoStop}%
\bibitem [{\citenamefont {Rodríguez}, \citenamefont {Sengupta},\ and\
  \citenamefont {Bhattacharjee}(2022)}]{rodriguez2022phase}%
  \BibitemOpen
  \bibfield  {author} {\bibinfo {author} {\bibfnamefont {E.}~\bibnamefont
  {Rodríguez}}, \bibinfo {author} {\bibfnamefont {W.}~\bibnamefont
  {Sengupta}},\ and\ \bibinfo {author} {\bibfnamefont {A.}~\bibnamefont
  {Bhattacharjee}},\ }\bibfield  {title} {\enquote {\bibinfo {title} {Phases
  and phase-transitions in quasisymmetric configuration space},}\ }\href
  {https://doi.org/10.1088/1361-6587/ac89af} {\bibfield  {journal} {\bibinfo
  {journal} {Plasma Physics and Controlled Fusion}\ }\textbf {\bibinfo {volume}
  {64}},\ \bibinfo {pages} {105006} (\bibinfo {year} {2022})}\BibitemShut
  {NoStop}%
\bibitem [{Note1()}]{Note1}%
  \BibitemOpen
  \bibinfo {note} {We simplify the picture by assuming that the electrostatic
  potential shares the QS to leading gyro-order and do not include it in our
  considerations.}\BibitemShut {Stop}%
\bibitem [{\citenamefont {Tessarotto}\ \emph {et~al.}(1996)\citenamefont
  {Tessarotto}, \citenamefont {Johnson}, \citenamefont {White},\ and\
  \citenamefont {Zheng}}]{tessarotto1996}%
  \BibitemOpen
  \bibfield  {author} {\bibinfo {author} {\bibfnamefont {M.}~\bibnamefont
  {Tessarotto}}, \bibinfo {author} {\bibfnamefont {J.~L.}\ \bibnamefont
  {Johnson}}, \bibinfo {author} {\bibfnamefont {R.~B.}\ \bibnamefont {White}},\
  and\ \bibinfo {author} {\bibfnamefont {L.}~\bibnamefont {Zheng}},\ }\bibfield
   {title} {\enquote {\bibinfo {title} {Quasi‐helical magnetohydrodynamic
  equilibria in the presence of flow},}\ }\href
  {https://doi.org/10.1063/1.871522} {\bibfield  {journal} {\bibinfo  {journal}
  {Physics of Plasmas}\ }\textbf {\bibinfo {volume} {3}},\ \bibinfo {pages}
  {2653--2663} (\bibinfo {year} {1996})}\BibitemShut {NoStop}%
\bibitem [{\citenamefont {Burby}, \citenamefont {Kallinikos},\ and\
  \citenamefont {MacKay}(2021)}]{burby2020}%
  \BibitemOpen
  \bibfield  {author} {\bibinfo {author} {\bibfnamefont {J.~W.}\ \bibnamefont
  {Burby}}, \bibinfo {author} {\bibfnamefont {N.}~\bibnamefont {Kallinikos}},\
  and\ \bibinfo {author} {\bibfnamefont {R.~S.}\ \bibnamefont {MacKay}},\
  }\bibfield  {title} {\enquote {\bibinfo {title} {Approximate symmetries of
  guiding-centre motion},}\ }\href {https://doi.org/10.1088/1751-8121/abe58a}
  {\bibfield  {journal} {\bibinfo  {journal} {Journal of Physics A:
  Mathematical and Theoretical}\ }\textbf {\bibinfo {volume} {54}},\ \bibinfo
  {pages} {125202} (\bibinfo {year} {2021})}\BibitemShut {NoStop}%
\bibitem [{\citenamefont {Rodríguez}, \citenamefont {Paul},\ and\
  \citenamefont {Bhattacharjee}(2022)}]{rodriguez2021opt}%
  \BibitemOpen
  \bibfield  {author} {\bibinfo {author} {\bibfnamefont {E.}~\bibnamefont
  {Rodríguez}}, \bibinfo {author} {\bibfnamefont {E.}~\bibnamefont {Paul}},\
  and\ \bibinfo {author} {\bibfnamefont {A.}~\bibnamefont {Bhattacharjee}},\
  }\bibfield  {title} {\enquote {\bibinfo {title} {Measures of quasisymmetry
  for stellarators},}\ }\href {https://doi.org/10.1017/S0022377821001331}
  {\bibfield  {journal} {\bibinfo  {journal} {Journal of Plasma Physics}\
  }\textbf {\bibinfo {volume} {88}},\ \bibinfo {pages} {905880109} (\bibinfo
  {year} {2022})}\BibitemShut {NoStop}%
\bibitem [{Note2()}]{Note2}%
  \BibitemOpen
  \bibinfo {note} {For a more general form of equilibrium, a formally analogous
  approach exists in terms of so-called generalised Boozer coordinates, details
  of which may be found in [\protect \rev@citealpnum {rodrigGBC}]. As a result,
  many of the properties of quasisymmetric stellarators in this paper are
  independent of a particular form of equilibrium.}\BibitemShut {Stop}%
\bibitem [{\citenamefont {Garren}\ and\ \citenamefont
  {Boozer}(1991{\natexlab{a}})}]{garrenboozer1991a}%
  \BibitemOpen
  \bibfield  {author} {\bibinfo {author} {\bibfnamefont {D.~A.}\ \bibnamefont
  {Garren}}\ and\ \bibinfo {author} {\bibfnamefont {A.~H.}\ \bibnamefont
  {Boozer}},\ }\bibfield  {title} {\enquote {\bibinfo {title} {Magnetic field
  strength of toroidal plasma equilibria},}\ }\href
  {https://doi.org/10.1063/1.859915} {\bibfield  {journal} {\bibinfo  {journal}
  {Physics of Fluids B: Plasma Physics}\ }\textbf {\bibinfo {volume} {3}},\
  \bibinfo {pages} {2805--2821} (\bibinfo {year}
  {1991}{\natexlab{a}})}\BibitemShut {NoStop}%
\bibitem [{\citenamefont {Landreman}, \citenamefont {Sengupta},\ and\
  \citenamefont {Plunk}(2019)}]{landreman2019}%
  \BibitemOpen
  \bibfield  {author} {\bibinfo {author} {\bibfnamefont {M.}~\bibnamefont
  {Landreman}}, \bibinfo {author} {\bibfnamefont {W.}~\bibnamefont
  {Sengupta}},\ and\ \bibinfo {author} {\bibfnamefont {G.~G.}\ \bibnamefont
  {Plunk}},\ }\bibfield  {title} {\enquote {\bibinfo {title} {Direct
  construction of optimized stellarator shapes. part 2. numerical
  quasisymmetric solutions},}\ }\href
  {https://doi.org/10.1017/S0022377818001344} {\bibfield  {journal} {\bibinfo
  {journal} {Journal of Plasma Physics}\ }\textbf {\bibinfo {volume} {85}},\
  \bibinfo {pages} {905850103} (\bibinfo {year} {2019})}\BibitemShut {NoStop}%
\bibitem [{\citenamefont {Rodríguez}\ and\ \citenamefont
  {Bhattacharjee}(2021)}]{rodriguez2020i}%
  \BibitemOpen
  \bibfield  {author} {\bibinfo {author} {\bibfnamefont {E.}~\bibnamefont
  {Rodríguez}}\ and\ \bibinfo {author} {\bibfnamefont {A.}~\bibnamefont
  {Bhattacharjee}},\ }\bibfield  {title} {\enquote {\bibinfo {title} {Solving
  the problem of overdetermination of quasisymmetric equilibrium solutions by
  near-axis expansions. i. generalized force balance},}\ }\href
  {https://doi.org/10.1063/5.0027574} {\bibfield  {journal} {\bibinfo
  {journal} {Physics of Plasmas}\ }\textbf {\bibinfo {volume} {28}},\ \bibinfo
  {pages} {012508} (\bibinfo {year} {2021})}\BibitemShut {NoStop}%
\bibitem [{\citenamefont {Kruskal}\ and\ \citenamefont
  {Kulsrud}(1958)}]{kruskuls1958}%
  \BibitemOpen
  \bibfield  {author} {\bibinfo {author} {\bibfnamefont {M.~D.}\ \bibnamefont
  {Kruskal}}\ and\ \bibinfo {author} {\bibfnamefont {R.~M.}\ \bibnamefont
  {Kulsrud}},\ }\bibfield  {title} {\enquote {\bibinfo {title} {Equilibrium of
  a magnetically confined plasma in a toroid},}\ }\href
  {https://doi.org/10.1063/1.1705884} {\bibfield  {journal} {\bibinfo
  {journal} {The Physics of Fluids}\ }\textbf {\bibinfo {volume} {1}},\
  \bibinfo {pages} {265--274} (\bibinfo {year} {1958})}\BibitemShut {NoStop}%
\bibitem [{\citenamefont {Rodríguez}, \citenamefont {Sengupta},\ and\
  \citenamefont {Bhattacharjee}(2021)}]{rodrigGBC}%
  \BibitemOpen
  \bibfield  {author} {\bibinfo {author} {\bibfnamefont {E.}~\bibnamefont
  {Rodríguez}}, \bibinfo {author} {\bibfnamefont {W.}~\bibnamefont
  {Sengupta}},\ and\ \bibinfo {author} {\bibfnamefont {A.}~\bibnamefont
  {Bhattacharjee}},\ }\bibfield  {title} {\enquote {\bibinfo {title}
  {Generalized boozer coordinates: A natural coordinate system for
  quasisymmetry},}\ }\href {https://doi.org/10.1063/5.0060115} {\bibfield
  {journal} {\bibinfo  {journal} {Physics of Plasmas}\ }\textbf {\bibinfo
  {volume} {28}},\ \bibinfo {pages} {092510} (\bibinfo {year}
  {2021})}\BibitemShut {NoStop}%
\bibitem [{\citenamefont {Mercier}(1964)}]{Mercier1964}%
  \BibitemOpen
  \bibfield  {author} {\bibinfo {author} {\bibfnamefont {C.}~\bibnamefont
  {Mercier}},\ }\bibfield  {title} {\enquote {\bibinfo {title} {{Equilibrium
  and stability of a toroidal magnetohydrodynamic system in the neighbourhood
  of a magnetic axis}},}\ }\href {https://doi.org/10.1088/0029-5515/4/3/008}
  {\bibfield  {journal} {\bibinfo  {journal} {Nuclear Fusion}\ }\textbf
  {\bibinfo {volume} {4}},\ \bibinfo {pages} {213} (\bibinfo {year}
  {1964})}\BibitemShut {NoStop}%
\bibitem [{\citenamefont {Solov'ev}\ and\ \citenamefont
  {Shafranov}(1970)}]{Solovev1970}%
  \BibitemOpen
  \bibfield  {author} {\bibinfo {author} {\bibfnamefont {L.~S.}\ \bibnamefont
  {Solov'ev}}\ and\ \bibinfo {author} {\bibfnamefont {V.~D.}\ \bibnamefont
  {Shafranov}},\ }\href@noop {} {\emph {\bibinfo {title} {{Reviews of Plasma
  Physics 5}}}}\ (\bibinfo  {publisher} {Consultants Bureau},\ \bibinfo
  {address} {New York - London},\ \bibinfo {year} {1970})\BibitemShut {NoStop}%
\bibitem [{Note3()}]{Note3}%
  \BibitemOpen
  \bibinfo {note} {For simplicity, we have not normalised $\psi $ respect to
  the magnetic field on the axis and its curvature as it is often
  customary\cite {garrenboozer1991a,landreman2019}. Doing so simply introduces
  constant rescaling factors in various quantities involved.}\BibitemShut
  {Stop}%
\bibitem [{\citenamefont {Garren}\ and\ \citenamefont
  {Boozer}(1991{\natexlab{b}})}]{garrenboozer1991b}%
  \BibitemOpen
  \bibfield  {author} {\bibinfo {author} {\bibfnamefont {D.~A.}\ \bibnamefont
  {Garren}}\ and\ \bibinfo {author} {\bibfnamefont {A.~H.}\ \bibnamefont
  {Boozer}},\ }\bibfield  {title} {\enquote {\bibinfo {title} {Existence of
  quasihelically symmetric stellarators},}\ }\href
  {https://doi.org/10.1063/1.859916} {\bibfield  {journal} {\bibinfo  {journal}
  {Physics of Fluids B: Plasma Physics}\ }\textbf {\bibinfo {volume} {3}},\
  \bibinfo {pages} {2822--2834} (\bibinfo {year}
  {1991}{\natexlab{b}})}\BibitemShut {NoStop}%
\bibitem [{\citenamefont {Rodr{\'\i}guez}, \citenamefont {Sengupta},\ and\
  \citenamefont {Bhattacharjee}(2022)}]{rodriguez2021weak}%
  \BibitemOpen
  \bibfield  {author} {\bibinfo {author} {\bibfnamefont {E.}~\bibnamefont
  {Rodr{\'\i}guez}}, \bibinfo {author} {\bibfnamefont {W.}~\bibnamefont
  {Sengupta}},\ and\ \bibinfo {author} {\bibfnamefont {A.}~\bibnamefont
  {Bhattacharjee}},\ }\bibfield  {title} {\enquote {\bibinfo {title} {Weakly
  quasisymmetric near-axis solutions to all orders},}\ }\href@noop {}
  {\bibfield  {journal} {\bibinfo  {journal} {Physics of Plasmas}\ }\textbf
  {\bibinfo {volume} {29}},\ \bibinfo {pages} {012507} (\bibinfo {year}
  {2022})}\BibitemShut {NoStop}%
\bibitem [{\citenamefont {Rodriguez}(2022)}]{thesis}%
  \BibitemOpen
  \bibfield  {author} {\bibinfo {author} {\bibfnamefont {E.}~\bibnamefont
  {Rodriguez}},\ }\emph {\bibinfo {title} {Quasisymmetry}},\ \href@noop {}
  {Ph.D. thesis} (\bibinfo {year} {2022})\BibitemShut {NoStop}%
\bibitem [{\citenamefont {Landreman}\ and\ \citenamefont
  {Sengupta}(2018)}]{landreman2018a}%
  \BibitemOpen
  \bibfield  {author} {\bibinfo {author} {\bibfnamefont {M.}~\bibnamefont
  {Landreman}}\ and\ \bibinfo {author} {\bibfnamefont {W.}~\bibnamefont
  {Sengupta}},\ }\bibfield  {title} {\enquote {\bibinfo {title} {Direct
  construction of optimized stellarator shapes. part 1. theory in cylindrical
  coordinates},}\ }\href {https://doi.org/10.1017/S0022377818001289} {\bibfield
   {journal} {\bibinfo  {journal} {Journal of Plasma Physics}\ }\textbf
  {\bibinfo {volume} {84}},\ \bibinfo {pages} {905840616} (\bibinfo {year}
  {2018})}\BibitemShut {NoStop}%
\bibitem [{\citenamefont {Plunk}, \citenamefont {Landreman},\ and\
  \citenamefont {Helander}(2019)}]{plunk2019}%
  \BibitemOpen
  \bibfield  {author} {\bibinfo {author} {\bibfnamefont {G.~G.}\ \bibnamefont
  {Plunk}}, \bibinfo {author} {\bibfnamefont {M.}~\bibnamefont {Landreman}},\
  and\ \bibinfo {author} {\bibfnamefont {P.}~\bibnamefont {Helander}},\
  }\bibfield  {title} {\enquote {\bibinfo {title} {Direct construction of
  optimized stellarator shapes. part 3. omnigenity near the magnetic axis},}\
  }\href {https://doi.org/10.1017/S002237781900062X} {\bibfield  {journal}
  {\bibinfo  {journal} {Journal of Plasma Physics}\ }\textbf {\bibinfo {volume}
  {85}},\ \bibinfo {pages} {905850602} (\bibinfo {year} {2019})}\BibitemShut
  {NoStop}%
\bibitem [{\citenamefont {Rodriguez}\ and\ \citenamefont
  {Plunk}(2023)}]{rodriguez2023qi}%
  \BibitemOpen
  \bibfield  {author} {\bibinfo {author} {\bibfnamefont {E.}~\bibnamefont
  {Rodriguez}}\ and\ \bibinfo {author} {\bibfnamefont {G.~G.}\ \bibnamefont
  {Plunk}},\ }\bibfield  {title} {\enquote {\bibinfo {title} {Higher order
  theory of quasi-isodynamicity near the magnetic axis of stellarators},}\
  }\href@noop {} {\bibfield  {journal} {\bibinfo  {journal} {arXiv preprint
  arXiv:2303.06038}\ } (\bibinfo {year} {2023})}\BibitemShut {NoStop}%
\bibitem [{\citenamefont {Oberti}\ and\ \citenamefont
  {Ricca}(2016)}]{oberti2016}%
  \BibitemOpen
  \bibfield  {author} {\bibinfo {author} {\bibfnamefont {C.}~\bibnamefont
  {Oberti}}\ and\ \bibinfo {author} {\bibfnamefont {R.~L.}\ \bibnamefont
  {Ricca}},\ }\bibfield  {title} {\enquote {\bibinfo {title} {On torus knots
  and unknots},}\ }\href {https://doi.org/10.1142/S021821651650036X} {\bibfield
   {journal} {\bibinfo  {journal} {Journal of Knot Theory and Its
  Ramifications}\ }\textbf {\bibinfo {volume} {25}},\ \bibinfo {pages}
  {1650036} (\bibinfo {year} {2016})}\BibitemShut {NoStop}%
\bibitem [{\citenamefont {Aicardi}(2000)}]{aicardi2000}%
  \BibitemOpen
  \bibfield  {author} {\bibinfo {author} {\bibfnamefont {F.}~\bibnamefont
  {Aicardi}},\ }\bibfield  {title} {\enquote {\bibinfo {title} {Self-linking of
  spatial curves without inflections and its applications},}\ }\href@noop {}
  {\bibfield  {journal} {\bibinfo  {journal} {Funct Anal Its Appl}\ }\textbf
  {\bibinfo {volume} {34}},\ \bibinfo {pages} {79--85} (\bibinfo {year}
  {2000})}\BibitemShut {NoStop}%
\bibitem [{\citenamefont {Fuller}(1999)}]{fuller1999}%
  \BibitemOpen
  \bibfield  {author} {\bibinfo {author} {\bibfnamefont {J.}~\bibnamefont
  {Fuller}, \bibfnamefont {Edgar~J.}},\ }\href@noop {} {\emph {\bibinfo {title}
  {The geometric and topological structure of holonomic knots}}}\ (\bibinfo
  {year} {1999})\ p.\ \bibinfo {pages} {112}\BibitemShut {NoStop}%
\bibitem [{\citenamefont {Moffatt}\ and\ \citenamefont
  {Ricca}(1992)}]{moffatt1992}%
  \BibitemOpen
  \bibfield  {author} {\bibinfo {author} {\bibfnamefont {H.~K.}\ \bibnamefont
  {Moffatt}}\ and\ \bibinfo {author} {\bibfnamefont {R.~L.}\ \bibnamefont
  {Ricca}},\ }\bibfield  {title} {\enquote {\bibinfo {title} {Helicity and the
  c\"{a}lug\"{a}reanu invariant},}\ }\href
  {https://doi.org/10.1098/rspa.1992.0159} {\bibfield  {journal} {\bibinfo
  {journal} {Proceedings of the Royal Society of London. Series A: Mathematical
  and Physical Sciences}\ }\textbf {\bibinfo {volume} {439}},\ \bibinfo {pages}
  {411--429} (\bibinfo {year} {1992})}\BibitemShut {NoStop}%
\bibitem [{\citenamefont {Fenchel}(1951)}]{fenchel1951}%
  \BibitemOpen
  \bibfield  {author} {\bibinfo {author} {\bibfnamefont {W.}~\bibnamefont
  {Fenchel}},\ }\bibfield  {title} {\enquote {\bibinfo {title} {On the
  differential geometry of closed space curves},}\ }\href@noop {} {\bibfield
  {journal} {\bibinfo  {journal} {Bulletin of the American Mathematical
  Society}\ }\textbf {\bibinfo {volume} {57}},\ \bibinfo {pages} {44--54}
  (\bibinfo {year} {1951})}\BibitemShut {NoStop}%
\bibitem [{\citenamefont {Paz-Soldan}(2020)}]{paz2020}%
  \BibitemOpen
  \bibfield  {author} {\bibinfo {author} {\bibfnamefont {C.}~\bibnamefont
  {Paz-Soldan}},\ }\bibfield  {title} {\enquote {\bibinfo {title} {Non-planar
  coil winding angle optimization for compatibility with non-insulated
  high-temperature superconducting magnets},}\ }\href@noop {} {\bibfield
  {journal} {\bibinfo  {journal} {Journal of Plasma Physics}\ }\textbf
  {\bibinfo {volume} {86}},\ \bibinfo {pages} {815860501} (\bibinfo {year}
  {2020})}\BibitemShut {NoStop}%
\bibitem [{\citenamefont {Lonigro}\ and\ \citenamefont
  {Zhu}(2022)}]{lonigro2022}%
  \BibitemOpen
  \bibfield  {author} {\bibinfo {author} {\bibfnamefont {N.}~\bibnamefont
  {Lonigro}}\ and\ \bibinfo {author} {\bibfnamefont {C.}~\bibnamefont {Zhu}},\
  }\bibfield  {title} {\enquote {\bibinfo {title} {Stellarator coil design
  using cubic splines for improved access on the outboard side},}\ }\href@noop
  {} {\bibfield  {journal} {\bibinfo  {journal} {Nuclear Fusion}\ }\textbf
  {\bibinfo {volume} {62}},\ \bibinfo {pages} {066009} (\bibinfo {year}
  {2022})}\BibitemShut {NoStop}%
\bibitem [{\citenamefont {Helander}\ and\ \citenamefont
  {Sigmar}(2005)}]{helander2005}%
  \BibitemOpen
  \bibfield  {author} {\bibinfo {author} {\bibfnamefont {P.}~\bibnamefont
  {Helander}}\ and\ \bibinfo {author} {\bibfnamefont {D.}~\bibnamefont
  {Sigmar}},\ }\href@noop {} {\emph {\bibinfo {title} {Collisional Transport in
  Magnetized Plasmas}}},\ Cambridge Monographs on Plasma Physics\ (\bibinfo
  {publisher} {Cambridge University Press},\ \bibinfo {year}
  {2005})\BibitemShut {NoStop}%
\bibitem [{\citenamefont {Ware}\ \emph {et~al.}(2006)\citenamefont {Ware},
  \citenamefont {Spong}, \citenamefont {Berry}, \citenamefont {Hirshman},\ and\
  \citenamefont {Lyon}}]{ware2006}%
  \BibitemOpen
  \bibfield  {author} {\bibinfo {author} {\bibfnamefont {A.~S.}\ \bibnamefont
  {Ware}}, \bibinfo {author} {\bibfnamefont {D.~A.}\ \bibnamefont {Spong}},
  \bibinfo {author} {\bibfnamefont {L.~A.}\ \bibnamefont {Berry}}, \bibinfo
  {author} {\bibfnamefont {S.~P.}\ \bibnamefont {Hirshman}},\ and\ \bibinfo
  {author} {\bibfnamefont {J.~F.}\ \bibnamefont {Lyon}},\ }\bibfield  {title}
  {\enquote {\bibinfo {title} {Bootstrap current in quasi-symmetric
  stellarators},}\ }\href@noop {} {\bibfield  {journal} {\bibinfo  {journal}
  {Fusion Science and Technology}\ }\textbf {\bibinfo {volume} {50}},\ \bibinfo
  {pages} {236--244} (\bibinfo {year} {2006})}\BibitemShut {NoStop}%
\bibitem [{\citenamefont {Rodriguez}(2023{\natexlab{a}})}]{rodriguez2023mhd}%
  \BibitemOpen
  \bibfield  {author} {\bibinfo {author} {\bibfnamefont {E.}~\bibnamefont
  {Rodriguez}},\ }\bibfield  {title} {\enquote {\bibinfo {title} {Mhd stability
  and the effects of shaping: a near-axis view for tokamaks and quasisymmetric
  stellarators},}\ }\href@noop {} {\bibfield  {journal} {\bibinfo  {journal}
  {arXiv preprint arXiv:2302.03359}\ } (\bibinfo {year}
  {2023}{\natexlab{a}})}\BibitemShut {NoStop}%
\bibitem [{\citenamefont {Landreman}\ and\ \citenamefont
  {Paul}(2022)}]{landreman2022}%
  \BibitemOpen
  \bibfield  {author} {\bibinfo {author} {\bibfnamefont {M.}~\bibnamefont
  {Landreman}}\ and\ \bibinfo {author} {\bibfnamefont {E.}~\bibnamefont
  {Paul}},\ }\bibfield  {title} {\enquote {\bibinfo {title} {Magnetic fields
  with precise quasisymmetry for plasma confinement},}\ }\href@noop {}
  {\bibfield  {journal} {\bibinfo  {journal} {Physical Review Letters}\
  }\textbf {\bibinfo {volume} {128}},\ \bibinfo {pages} {035001} (\bibinfo
  {year} {2022})}\BibitemShut {NoStop}%
\bibitem [{Note4()}]{Note4}%
  \BibitemOpen
  \bibinfo {note} {From the Mercier perspective on rotational transform\cite
  {Helander2014}, $$ \protect \bar {\iota }_0=-\protect \frac {1}{2\pi }\DOTSI
  \intop \ilimits@ _0^L\protect \frac {\protect \qopname \relax o{cosh}\protect
  \bar {\eta }-1}{\protect \qopname \relax o{cosh}\protect \bar {\eta
  }}(d'+\tau )\protect \mathrm {d}l+\protect \frac {1}{2\pi }\DOTSI \intop
  \ilimits@ _0^L\tau \protect \mathrm {d}l, \label {eqn:MercierIota} $$ in the
  large ellipticity $\protect \bar {\eta }\rightarrow \infty $ limit, $\protect
  \bar {\iota }_0=-d/2\pi $, where $d$ is the angle of rotation of the ellipse
  with respect to the curvature vector. Because $\sigma \sim \eta ^2\rightarrow
  0$ in this limit, the cross-sections align with the Frenet-Serret frame.
  Thus, the net rotation $d=0$.}\BibitemShut {Stop}%
\bibitem [{\citenamefont {Drevlak}\ \emph {et~al.}(2013)\citenamefont
  {Drevlak}, \citenamefont {Brochard}, \citenamefont {Helander}, \citenamefont
  {Kisslinger}, \citenamefont {Mikhailov}, \citenamefont {N{\"u}hrenberg},
  \citenamefont {N{\"u}hrenberg},\ and\ \citenamefont {Turkin}}]{Drevlak2013}%
  \BibitemOpen
  \bibfield  {author} {\bibinfo {author} {\bibfnamefont {M.}~\bibnamefont
  {Drevlak}}, \bibinfo {author} {\bibfnamefont {F.}~\bibnamefont {Brochard}},
  \bibinfo {author} {\bibfnamefont {P.}~\bibnamefont {Helander}}, \bibinfo
  {author} {\bibfnamefont {J.}~\bibnamefont {Kisslinger}}, \bibinfo {author}
  {\bibfnamefont {M.}~\bibnamefont {Mikhailov}}, \bibinfo {author}
  {\bibfnamefont {C.}~\bibnamefont {N{\"u}hrenberg}}, \bibinfo {author}
  {\bibfnamefont {J.}~\bibnamefont {N{\"u}hrenberg}},\ and\ \bibinfo {author}
  {\bibfnamefont {Y.}~\bibnamefont {Turkin}},\ }\bibfield  {title} {\enquote
  {\bibinfo {title} {{ESTELL}: A quasi-toroidally symmetric stellarator},}\
  }\href@noop {} {\bibfield  {journal} {\bibinfo  {journal} {Contributions to
  Plasma Physics}\ }\textbf {\bibinfo {volume} {53}},\ \bibinfo {pages}
  {459--468} (\bibinfo {year} {2013})}\BibitemShut {NoStop}%
\bibitem [{\citenamefont {Garabedian}(2008)}]{Garabedian2008}%
  \BibitemOpen
  \bibfield  {author} {\bibinfo {author} {\bibfnamefont {P.~R.}\ \bibnamefont
  {Garabedian}},\ }\bibfield  {title} {\enquote {\bibinfo {title}
  {Three-dimensional analysis of tokamaks and stellarators},}\ }\href@noop {}
  {\bibfield  {journal} {\bibinfo  {journal} {Proceedings of the National
  Academy of Sciences}\ }\textbf {\bibinfo {volume} {105}},\ \bibinfo {pages}
  {13716--13719} (\bibinfo {year} {2008})}\BibitemShut {NoStop}%
\bibitem [{\citenamefont {Garabedian}\ and\ \citenamefont
  {McFadden}(2009)}]{Garabedian2009}%
  \BibitemOpen
  \bibfield  {author} {\bibinfo {author} {\bibfnamefont {P.~R.}\ \bibnamefont
  {Garabedian}}\ and\ \bibinfo {author} {\bibfnamefont {G.~B.}\ \bibnamefont
  {McFadden}},\ }\bibfield  {title} {\enquote {\bibinfo {title} {Design of the
  {DEMO} fusion reactor following {ITER}},}\ }\href@noop {} {\bibfield
  {journal} {\bibinfo  {journal} {Journal of research of the National Institute
  of Standards and Technology}\ }\textbf {\bibinfo {volume} {114}},\ \bibinfo
  {pages} {229} (\bibinfo {year} {2009})}\BibitemShut {NoStop}%
\bibitem [{\citenamefont {Hirshman}\ and\ \citenamefont
  {Whitson}(1983)}]{hirshman1983}%
  \BibitemOpen
  \bibfield  {author} {\bibinfo {author} {\bibfnamefont {S.~P.}\ \bibnamefont
  {Hirshman}}\ and\ \bibinfo {author} {\bibfnamefont {J.~C.}\ \bibnamefont
  {Whitson}},\ }\bibfield  {title} {\enquote {\bibinfo {title}
  {Steepest‐descent moment method for three‐dimensional magnetohydrodynamic
  equilibria},}\ }\href@noop {} {\bibfield  {journal} {\bibinfo  {journal} {The
  Physics of Fluids}\ }\textbf {\bibinfo {volume} {26}},\ \bibinfo {pages}
  {3553--3568} (\bibinfo {year} {1983})}\BibitemShut {NoStop}%
\bibitem [{\citenamefont {Landreman}(2021)}]{landreman2021a}%
  \BibitemOpen
  \bibfield  {author} {\bibinfo {author} {\bibfnamefont {M.}~\bibnamefont
  {Landreman}},\ }\bibfield  {title} {\enquote {\bibinfo {title} {Figures of
  merit for stellarators near the magnetic axis},}\ }\href
  {https://doi.org/10.1017/S0022377820001658} {\bibfield  {journal} {\bibinfo
  {journal} {Journal of Plasma Physics}\ }\textbf {\bibinfo {volume} {87}},\
  \bibinfo {pages} {905870112} (\bibinfo {year} {2021})}\BibitemShut {NoStop}%
\bibitem [{\citenamefont {Landreman}(2022)}]{landreman2022map}%
  \BibitemOpen
  \bibfield  {author} {\bibinfo {author} {\bibfnamefont {M.}~\bibnamefont
  {Landreman}},\ }\bibfield  {title} {\enquote {\bibinfo {title} {Mapping the
  space of quasisymmetric stellarators using optimized near-axis expansion},}\
  }\href {https://doi.org/10.1017/S0022377822001258} {\bibfield  {journal}
  {\bibinfo  {journal} {Journal of Plasma Physics}\ }\textbf {\bibinfo {volume}
  {88}},\ \bibinfo {pages} {905880616} (\bibinfo {year} {2022})}\BibitemShut
  {NoStop}%
\bibitem [{\citenamefont {Bender}\ and\ \citenamefont
  {Orszag}(1999)}]{bender1999}%
  \BibitemOpen
  \bibfield  {author} {\bibinfo {author} {\bibfnamefont {C.~M.}\ \bibnamefont
  {Bender}}\ and\ \bibinfo {author} {\bibfnamefont {S.~A.}\ \bibnamefont
  {Orszag}},\ }\href@noop {} {\emph {\bibinfo {title} {Advanced mathematical
  methods for scientists and engineers: I: Asymptotic methods and perturbation
  theory}}}\ (\bibinfo  {publisher} {Springer},\ \bibinfo {year}
  {1999})\BibitemShut {NoStop}%
\bibitem [{\citenamefont {Landreman}\ and\ \citenamefont
  {Jorge}(2020)}]{landreman2020}%
  \BibitemOpen
  \bibfield  {author} {\bibinfo {author} {\bibfnamefont {M.}~\bibnamefont
  {Landreman}}\ and\ \bibinfo {author} {\bibfnamefont {R.}~\bibnamefont
  {Jorge}},\ }\bibfield  {title} {\enquote {\bibinfo {title} {Magnetic well and
  mercier stability of stellarators near the magnetic axis},}\ }\href
  {https://doi.org/10.1017/S002237782000121X} {\bibfield  {journal} {\bibinfo
  {journal} {Journal of Plasma Physics}\ }\textbf {\bibinfo {volume} {86}},\
  \bibinfo {pages} {905860510} (\bibinfo {year} {2020})}\BibitemShut {NoStop}%
\bibitem [{\citenamefont {Rodriguez}(2023{\natexlab{b}})}]{rodriguez2022mhd}%
  \BibitemOpen
  \bibfield  {author} {\bibinfo {author} {\bibfnamefont {E.}~\bibnamefont
  {Rodriguez}},\ }\bibfield  {title} {\enquote {\bibinfo {title} {Mhd stability
  and the effects of shaping: a near-axis view for tokamaks and quasisymmetric
  stellarators},}\ }\href@noop {} {\bibfield  {journal} {\bibinfo  {journal}
  {arXiv preprint arXiv:2302.03359}\ } (\bibinfo {year}
  {2023}{\natexlab{b}})}\BibitemShut {NoStop}%
\bibitem [{\citenamefont {Wright}, \citenamefont {Nocedal}\ \emph
  {et~al.}(1999)\citenamefont {Wright}, \citenamefont {Nocedal} \emph
  {et~al.}}]{wright1999}%
  \BibitemOpen
  \bibfield  {author} {\bibinfo {author} {\bibfnamefont {S.}~\bibnamefont
  {Wright}}, \bibinfo {author} {\bibfnamefont {J.}~\bibnamefont {Nocedal}},
  \emph {et~al.},\ }\bibfield  {title} {\enquote {\bibinfo {title} {Numerical
  optimization},}\ }\href@noop {} {\bibfield  {journal} {\bibinfo  {journal}
  {Springer Science}\ }\textbf {\bibinfo {volume} {35}},\ \bibinfo {pages} {7}
  (\bibinfo {year} {1999})}\BibitemShut {NoStop}%
\bibitem [{Note5()}]{Note5}%
  \BibitemOpen
  \bibinfo {note} {See https://github.com/landreman/qsc. The script used to
  obtain the main plot in Fig.~\ref {fig:spaceQSquality} can be found in the
  Zenodo repository associated to this paper. The same may be achieved, albeit
  slower, using \protect \texttt {pyQSC}, which was how it was originally done
  and is also included there.}\BibitemShut {Stop}%
\bibitem [{Note6()}]{Note6}%
  \BibitemOpen
  \bibinfo {note} {An 11th Gen i7-11850H core was used for this purpose. The
  main space in Fig.~\ref {fig:spaceQSquality} (which is 300x300) took a total
  of about 14~hrs. Most time is devoted to the optimisation sub-problems at
  each point (search for $\eta ^*$, $B_{22}^C$, and $\{Z_n\}$). Of course, the
  construction of such a space is trivially parallelisable.}\BibitemShut
  {Stop}%
\bibitem [{\citenamefont {Plunk}\ and\ \citenamefont
  {Helander}(2018)}]{plunk2018}%
  \BibitemOpen
  \bibfield  {author} {\bibinfo {author} {\bibfnamefont {G.}~\bibnamefont
  {Plunk}}\ and\ \bibinfo {author} {\bibfnamefont {P.}~\bibnamefont
  {Helander}},\ }\bibfield  {title} {\enquote {\bibinfo {title}
  {Quasi-axisymmetric magnetic fields: weakly non-axisymmetric case in a
  vacuum},}\ }\href@noop {} {\bibfield  {journal} {\bibinfo  {journal} {Journal
  of Plasma Physics}\ }\textbf {\bibinfo {volume} {84}},\ \bibinfo {pages}
  {905840205} (\bibinfo {year} {2018})}\BibitemShut {NoStop}%
\bibitem [{\citenamefont {Nemov}\ \emph {et~al.}(1999)\citenamefont {Nemov},
  \citenamefont {Kasilov}, \citenamefont {Kernbichler},\ and\ \citenamefont
  {Heyn}}]{nemov1999}%
  \BibitemOpen
  \bibfield  {author} {\bibinfo {author} {\bibfnamefont {V.~V.}\ \bibnamefont
  {Nemov}}, \bibinfo {author} {\bibfnamefont {S.~V.}\ \bibnamefont {Kasilov}},
  \bibinfo {author} {\bibfnamefont {W.}~\bibnamefont {Kernbichler}},\ and\
  \bibinfo {author} {\bibfnamefont {M.~F.}\ \bibnamefont {Heyn}},\ }\bibfield
  {title} {\enquote {\bibinfo {title} {Evaluation of $1/\nu$ neoclassical
  transport in stellarators},}\ }\href {https://doi.org/10.1063/1.873749}
  {\bibfield  {journal} {\bibinfo  {journal} {Physics of Plasmas}\ }\textbf
  {\bibinfo {volume} {6}},\ \bibinfo {pages} {4622--4632} (\bibinfo {year}
  {1999})}\BibitemShut {NoStop}%
\bibitem [{\citenamefont {Nemov}, \citenamefont {Kasilov},\ and\ \citenamefont
  {Kernbichler}(2014)}]{nemov2014}%
  \BibitemOpen
  \bibfield  {author} {\bibinfo {author} {\bibfnamefont {V.~V.}\ \bibnamefont
  {Nemov}}, \bibinfo {author} {\bibfnamefont {S.~V.}\ \bibnamefont {Kasilov}},\
  and\ \bibinfo {author} {\bibfnamefont {W.}~\bibnamefont {Kernbichler}},\
  }\bibfield  {title} {\enquote {\bibinfo {title} {Collisionless high energy
  particle losses in optimized stellarators calculated in real-space
  coordinates},}\ }\href {https://doi.org/10.1063/1.4876740} {\bibfield
  {journal} {\bibinfo  {journal} {Physics of Plasmas}\ }\textbf {\bibinfo
  {volume} {21}},\ \bibinfo {pages} {062501} (\bibinfo {year}
  {2014})}\BibitemShut {NoStop}%
\bibitem [{\citenamefont {Wesson}(2011)}]{wessonTok}%
  \BibitemOpen
  \bibfield  {author} {\bibinfo {author} {\bibfnamefont {J.}~\bibnamefont
  {Wesson}},\ }\href@noop {} {\emph {\bibinfo {title} {{Tokamaks; 4th ed.}}}},\
  International series of monographs on physics\ (\bibinfo  {publisher} {Oxford
  Univ. Press},\ \bibinfo {address} {Oxford},\ \bibinfo {year}
  {2011})\BibitemShut {NoStop}%
\bibitem [{\citenamefont {Jorge}\ \emph {et~al.}(2022)\citenamefont {Jorge},
  \citenamefont {Plunk}, \citenamefont {Drevlak}, \citenamefont {Landreman},
  \citenamefont {Lobsien}, \citenamefont {Camacho~Mata},\ and\ \citenamefont
  {Helander}}]{jorge2022}%
  \BibitemOpen
  \bibfield  {author} {\bibinfo {author} {\bibfnamefont {R.}~\bibnamefont
  {Jorge}}, \bibinfo {author} {\bibfnamefont {G.}~\bibnamefont {Plunk}},
  \bibinfo {author} {\bibfnamefont {M.}~\bibnamefont {Drevlak}}, \bibinfo
  {author} {\bibfnamefont {M.}~\bibnamefont {Landreman}}, \bibinfo {author}
  {\bibfnamefont {J.-F.}\ \bibnamefont {Lobsien}}, \bibinfo {author}
  {\bibfnamefont {K.}~\bibnamefont {Camacho~Mata}},\ and\ \bibinfo {author}
  {\bibfnamefont {P.}~\bibnamefont {Helander}},\ }\bibfield  {title} {\enquote
  {\bibinfo {title} {A single-field-period quasi-isodynamic stellarator},}\
  }\href {https://doi.org/10.1017/S0022377822000873} {\bibfield  {journal}
  {\bibinfo  {journal} {Journal of Plasma Physics}\ }\textbf {\bibinfo {volume}
  {88}},\ \bibinfo {pages} {175880504} (\bibinfo {year} {2022})}\BibitemShut
  {NoStop}%
\bibitem [{\citenamefont {Griffiths}(2004)}]{griffiths2004QM}%
  \BibitemOpen
  \bibfield  {author} {\bibinfo {author} {\bibfnamefont {D.~J.}\ \bibnamefont
  {Griffiths}},\ }\href@noop {} {\emph {\bibinfo {title} {{Introduction to
  Quantum Mechanics (2nd Edition)}}}},\ \bibinfo {edition} {2nd}\ ed.\
  (\bibinfo  {publisher} {Pearson Prentice Hall},\ \bibinfo {year}
  {2004})\BibitemShut {NoStop}%
\end{thebibliography}%

\end{document}